\documentclass[iop]{emulateapj}

\usepackage{graphicx}	% Including figure files
\usepackage{amsmath}	% Advanced maths commands

\usepackage{amssymb}	% Extra maths symbols
\usepackage{bm}
\usepackage{booktabs}
\usepackage{cleveref}
\usepackage{natbib}
\usepackage{threeparttable}
\usepackage{xcolor}
\newcommand{\secpoint}{\mbox{$''\mskip-7.6mu.\,$}}

\begin{document}

\title{The ALMA REBELS Survey:  Specific Star-Formation Rates in the Reionization Era}

\author{
Michael W. Topping\altaffilmark{1},
Daniel P. Stark\altaffilmark{1},
Ryan Endsley\altaffilmark{1},
Rychard J. Bouwens\altaffilmark{2},
Sander Schouws\altaffilmark{2},
Renske Smit\altaffilmark{3},
Mauro Stefanon\altaffilmark{2},
Hanae Inami\altaffilmark{4},
Rebecca A. A. Bowler\altaffilmark{5},
Pascal Oesch\altaffilmark{6,7},
Valentino Gonzalez\altaffilmark{8,9}, 
Pratika Dayal\altaffilmark{10},
Elisabete da Cunha\altaffilmark{11},
Hiddo Algera\altaffilmark{4,12},
Paul van der Werf\altaffilmark{2},
Andrea Pallottini\altaffilmark{13},
Laia Barrufet De Soto\altaffilmark{7},
Raffaella Schneider\altaffilmark{14,15,16,17},
Ilse De Looze\altaffilmark{18,19},
Laura Sommovigo\altaffilmark{13},
Lily Whitler\altaffilmark{1},
Luca Graziani\altaffilmark{14,20},
Yoshinobu Fudamoto\altaffilmark{12,21},
Andrea Ferrara\altaffilmark{13}
}

\altaffiltext{1}{Steward Observatory, University of Arizona, 933 N Cherry Ave, Tucson, AZ 85721, USA}
\altaffiltext{2}{Leiden Observatory, Leiden University, NL-2300 RA Leiden, Netherlands}
\altaffiltext{3}{Astrophysics Research Institute, Liverpool John Moores University, 146 Brownlow Hill, Liverpool L3 5RF, United Kingdom}
\altaffiltext{4}{Hiroshima Astrophysical Science Center, Hiroshima University, 1-3-1 Kagamiyama, Higashi-Hiroshima, Hiroshima 739-8526, Japan}
%\altaffiltext{5}{Astrophysics, The Denys Wilkinson Building, University of Oxford, Keble Road, Oxford, OX1 3RH, United Kingdom}
\altaffiltext{5}{Jodrell Bank Centre for Astrophysics, Department of Physics and Astronomy, School of Natural Sciences, The University of Manchester, 
\,\, Manchester, M13 9PL, UK}
\altaffiltext{6}{Observatoire de Gen\'{e}ve, 1290 Versoix, Switzerland}
\altaffiltext{7}{Cosmic Dawn Center (DAWN), Niels Bohr Institute, University of Copenhagen, Jagtvej 128, K{\o}benhavn N, DK-2200, Denmark}
\altaffiltext{8}{Departmento de Astronomia, Universidad de Chile, Casilla 36-D, Santiago 7591245, Chile}
\altaffiltext{9}{Centro de Astrofisica y Tecnologias Afines (CATA), Camino del Observatorio 1515, Las Condes, Santiago, 7591245, Chile}
\altaffiltext{10}{Kapteyn Astronomical Institute, University of Groningen, PO Box 800, NL-9700 AV Groningen, the Netherlands}
\altaffiltext{11}{International Centre for Radio Astronomy Research, University of Western Australia, 35 Stirling Hwy., Crawley, WA 6009, Australia}
\altaffiltext{12}{National Astronomical Observatory of Japan, 2-21-1, Osawa, Mitaka, Tokyo, Japan}
\altaffiltext{13}{Scuola Normale Superiore, Piazza dei Cavalieri 7, 50126 Pisa, Italy}
\altaffiltext{14}{Dipartimento di Fisica, Sapienza, Universita di Roma, Piazzale Aldo Moro 5, I-00185 Roma, Italy}
\altaffiltext{15}{INAF/Osservatorio Astronomico di Roma, via Frascati 33, 00078 Monte Porzio Catone, Roma, Italy}
\altaffiltext{16}{Sapienza School for Advanced Studies, Sapienza Universit\`{a} di Roma, P.le Aldo Moro 2, 00185 Roma, Italy}
\altaffiltext{17}{INFN, Sezione di Roma 1, P.le Aldo Moro 2, 00185 Roma, Italy}
\altaffiltext{18}{Sterrenkundig Observatorium, Ghent University, Krijgslaan 281 - S9, 9000 Gent, Belgium}
\altaffiltext{19}{Dept. of Physics \& Astronomy, University College London, Gower Street, London WC1E 6BT, United Kingdom}
\altaffiltext{20}{INAF/Osservatorio Astrofisico di Arcetri, Largo E. Femi 5, I-50125 Firenze, Italy}
\altaffiltext{21}{Department of Physics, School of Advanced Science and Engineering, Faculty of Science and Engineering, Waseda University, 3-4-1, Okubo, Shinjuku, Tokyo 169-8555, Japan}

\email{michaeltopping@arizona.edu}

\shortauthors{Topping et al.}

\shorttitle{REBELS: sSFR at $z\sim7$}

%\label{firstpage}
%\maketitle

\begin{abstract}
We present specific star-formation rates (sSFRs) for 40 UV-bright galaxies at $z\sim7-8$ observed as part of the Reionization Era Bright Emission Line Survey (REBELS) ALMA large program. The sSFRs are derived using improved measures of SFR and stellar masses, made possible by  measurements of far-infrared (FIR) continuum emission and [CII]-based spectroscopic redshifts. For each source in the sample, we derive stellar masses from SED fitting and total SFRs from calibrations of the UV and FIR emission. The median sSFR is $18_{-5}^{+7}$ Gyr$^{-1}$, significantly larger than literature measurements lacking constraints in the FIR. The increase in sSFR reflects the larger obscured star formation rates we derive from the dust continuum relative to that implied by the UV+optical SED. We suggest that such differences may reflect spatial variations in dust across these luminous  galaxies, with the component dominating the FIR  distinct  from that dominating the UV. We demonstrate that the inferred stellar masses (and hence sSFRs) are strongly-dependent on the assumed star formation history in reionization-era galaxies. When large sSFR 
galaxies (a population which is common at $z>6$) are 
modeled with non-parametric star formation histories, 
the derived stellar masses can increase by an order of magnitude relative to constant star formation models, owing to the presence of a significant old stellar population that is outshined by the recent burst. The [CII] line widths in the largest sSFR systems are 
often very broad, suggesting dynamical masses that are easily able to accommodate the dominant old stellar population suggested by non-parametric models. Regardless of these systematic uncertainties in the derived parameters, we find  that the sSFR increases rapidly toward higher redshifts for massive galaxies ($9.6 < \log(\rm M_*/M_{\odot}) < 9.8$), with a power law that goes as $(1+z)^{1.7\pm0.3}$, broadly consistent with that expected from the evolving baryon accretion rates. 

\end{abstract}


%=============================================
%
%         INTRODUCTION
%
%==============================================
\section{Introduction} 
\label{sec:intro}
Deep imaging surveys using large ground and space-based telescopes in the past decade have revealed a wealth of information about galaxies in the epoch of reionization (see \citealt{Robertson2021} for a review). These observations have revealed an abundant population of 
relatively low luminosity star forming systems that likely contribute greatly to the ionizing budget required for reionization \citep[e.g.,][]{Bouwens2015, Finkelstein2015, Robertson2015, Ishigaki2018, Oesch2018, Naidu2021}. Much has 
been learned about the properties of early galaxies from the  rest-UV and optical spectral energy distributions (SEDs) constructed from the combination of {\it Hubble} and {\it Spitzer} photometry. The star formation rates (SFRs) and 
stellar masses implied by these SEDs allow for a variety of constraints on measures of galaxy growth through the reionization era \citep[e.g.,][]{Smit2016, Song2016, Stefanon2021}.

The specific SFR (sSFR$\equiv\rm SFR/M_*$) is one of the most useful measures of galaxy stellar mass build-up. When considering galaxies of fixed mass, the sSFR is generally predicted to increase with redshift, driven by the rise in baryon accretion rates at earlier times  \citep{Dekel2009, Fakhouri2010, Weinmann2011, Dave2011, Dayal2013, Krumholz2013, Correa2015, Sparre2015}. These theoretical expectations suggest the redshift evolution of the sSFR should follow a power law roughly of the form 
sSFR $\propto (1+z)^{2.25}$ \citep[e.g.,][]{Dekel2009}.  Deviations from this evolutionary form could arise for a variety of reasons if the SFRs of early galaxies are unable 
to keep up with the rapidly inflowing rate of baryons \citep[e.g.,][]{Gabor2014}.

Efforts to observationally constrain the redshift evolution of the sSFR into the reionization era began over a decade ago following the first {\it Hubble } and {\it Spitzer} deep fields.
Early results revealed similar sSFRs in galaxies of 
fixed mass at $2<z<7$. This suggested little evolution at redshifts higher than $z=2$ \citep[e.g.,][]{Stark2009, Labbe2010,Gonzalez2010, Gonzalez2011, Bouwens2012a}, in 
conflict with the simple predictions from the evolving 
baryon accretion rates, \citep[e.g.,][]{Weinmann2011}. 
As data and models improved, it became clear that the stellar masses at $z>5$ needed to be revised downward owing to a significant contribution from nebular emission lines in the {\it Spitzer}/IRAC bandpasses \citep{Schaerer2009}.  
Once accounted for, the sSFRs in the reionization era 
were found to be significantly larger than initial estimates suggested  \citep{Stark2013, Duncan2014, Gonzalez2014,Smit2014,Tasca2015}, easing tension with 
the redshift evolution predicted from rising baryon accretion rates. 

The most recent updates to the $z>4$ sSFRs have 
come from ALMA measurements of the thermal dust continuum in the far-infrared (FIR), providing a more direct constraint on obscured star formation in early galaxies. The ALPINE survey \citep{LeFevre2020, Bethermin2020, Faisst2020} presented the first statistical view of the dust continuum emission in $z\simeq 4.4-5.9$ UV-selected galaxies.  This enables much improved measurement of the total SFRs, through the combination of UV (unobscured) and FIR (obscured) calibrations. Using the derived UV+IR SFRs and stellar masses from ALPINE, \citet{Khusanova2021} characterized the average sSFR 
evolution.  The results suggested very slow evolution at $z>4$,  potentially again suggesting divergence from 
the rapid rise in sSFR predicted from the rising baryon accretion rates.

Here we extend this work into the reionization era using the 
sample of 40 UV-bright ($\rm M_{\rm UV} \lesssim -21.5$) galaxies at $z\sim7-9$ observed as part of the ALMA 
Reionization Era Bright Emission Line Survey \citep[REBELS;][]{Bouwens2021}. This sample marks a significant increase in the number of spectroscopic redshifts (via [CII] emission) and dust continuum detections in the reionization era. We use these data to characterise the sSFRs of UV-bright galaxies at this crucial epoch, for which our goals are twofold. First, we aim to explore the redshift evolution of the sSFR, using the improved constraints on the obscured SFR made possible 
by the ALMA continuum measurements. Second, we explore what the ALMA measurements reveal about the nature of the largest sSFR galaxies, a population 
of recent bursts that may contribute significantly to reionization 
\citep[e.g.,][]{Izotov2018, Tang2019, Vanzella2021, Izotov2021, Endsley2021, Naidu2022}. In Section 2 we provide an overview of the survey and observations. Section 3 describes the derivation of galaxy properties and calculation of the sSFRs.  Section 4 presents our main results with further discussion in Section 5. Finally, we provide a summary in Section 6.  Throughout this paper we assume a cosmology with $H_0=70~\rm km/s/Mpc$, $\Omega_{m}=0.30$, and $\Omega_{\Lambda}=0.70$.

%=============================================
%
%         METHODS
%
%==============================================
\section{Data and Methods}
\label{sec:data}
\subsection{The REBELS Survey}
The REBELS survey was designed to construct the first measurements of ISM cooling lines and dust continua for a statistical sample of UV-bright galaxies photometrically selected at $z>6.5$. A detailed description of the sample selection is provided in \citet{Bouwens2021}, however we provide a brief description here.  Candidate objects were selected in a number of fields with coverage in the optical, NIR, and Spitzer/IRAC bands including COSMOS/UltraVISTA, VIDEO/XMM-LSS+UKIDSS/UDS, and HST legacy fields, in addition to the BoRG/HIPPIES pure parallel fields \citep{Lawrence2007, Grogin2011, Koekemoer2011, Trenti2011,Yan2011, Mauduit2012, McCracken2012, Bradley2012, Postman2012, Jarvis2013, Schmidt2014,Steinhardt2014, Ashby2018, Coe2019, Salmon2020, Morishita2020, Roberts-Borsani2021}.  The candidate sample was narrowed down to a collection of UV-bright galaxies with constrained photometric redshifts selected from the source catalogs of \citet{Bowler2014, Bowler2017, Stefanon2017, Stefanon2019, Bowler2020, Endsley2021, Schouws2021, Bouwens2021, Stefanon2021inprep}.  The final targeted sample was then constructed of galaxies for which an ISM cooling line would likely be detected, which was determined using the measured UV luminosity converted to line flux using the calibration of \citet{deLooze2014}. This observed sample comprises 40 galaxies targeted within the redshift range $z=6.5-9.4$.

\begin{figure*}
    \centering
    \includegraphics[width=1.0\linewidth]{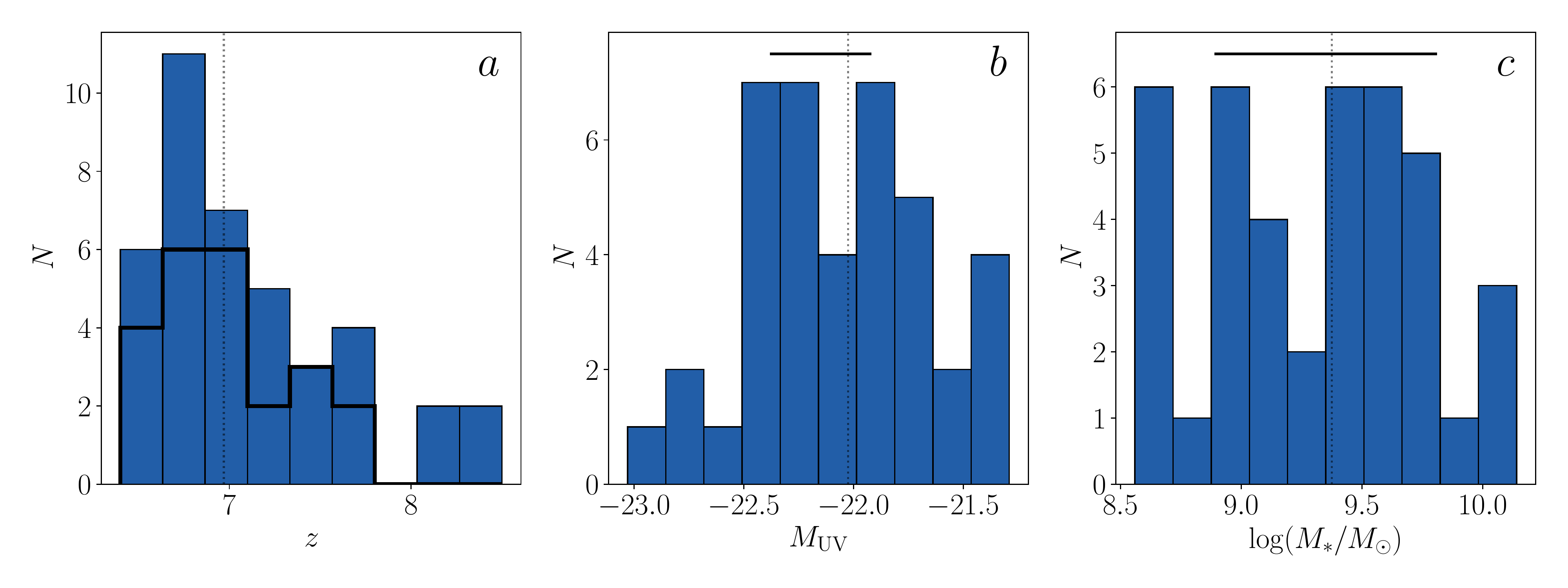}
    \caption{Sample properties of objects in the REBELS sample. The vertical dotted lines indicate the median value for each quantity. {\it a:} Redshift distribution of objects in the REBELS sample (blue histogram), where best-fit photometric redshifts are used for galaxies without spectroscopic redshifts. Redshift distribution of only objects with spectroscopic measurements is indicated by the black histogram. {\it b:} The sample $M_{\rm UV}$ of REBELS sources. This distribution is described by a median value of $-22.0$, with the sample ranging from $-21.3$ to $-23.0$. The typical $M_{\rm UV}$ error for individual measurements is indicated by the horizontal black line. {\it c:} Stellar mass distribution of REBELS galaxies inferred using the BEAGLE SED fitting code \citep{Chevallard2016} and assuming a CSFH (as described in Section~\ref{sec:masses}). This sample spans stellar masses of $\log(M_*/M_{\odot})=8.56-10.14$ with a median value of $\log(M_*/M_{\odot})=9.38$. The typical stellar mass error for individual measurements is indicated by the horizontal black line.} 
    \label{fig:sampleproperties}
\end{figure*}

Figure~\ref{fig:sampleproperties} shows the redshift, $\rm M_{\rm UV}$, and stellar mass distributions for the final targeted sample.  The redshift distribution, which is characterized by a median of $z=6.96$, comprises redshifts measured spectroscopically for 25 objects and the best-fit photometric redshift if no emission line could be measured. The REBELS sample spans absolute UV magnitudes in the range $-21.3$ to $-23.0$ with a median value of $\rm M_{\rm UV}=-22.0$.  This distribution is comparable to the ALPINE sample, which probes  $ \rm M_{\rm UV}=-20.2$ to $-22.7$ \citep{Faisst2020}. Finally, Figure~\ref{fig:sampleproperties}(c) shows the stellar mass distribution derived using SED fitting described below. This stellar mass distribution spans a similar range of stellar masses to that of the ALPINE survey \citep{Faisst2020}.  The similarity in $\rm M_{\rm UV}$ and stellar masses between the REBELS and ALPINE galaxies makes it possible to compare the two 
samples with the goal of understanding evolution of properties from $z\sim4.5$ to $z\sim7$.

\subsection{Observations and Data Reductions}
\label{sec:obs}
Observations of [CII]158$\mu$m, [OIII]88$\mu$m, and dust continua for the REBELS sources were obtained using ALMA. These observations consists of scans of spectral windows that cover the allowed observed frequency range of targeted ISM cooling lines determined by the photometric redshift likelihood distribution. The scans for emission lines in REBELS targets achieved the sensitivity required to detect [CII] of $2\times10^8 L_{\odot}$ at $5\sigma$ for a galaxy at $z=7$, and assuming a typical line width of $250\rm km/s$ \citep{Bouwens2021}.  For greatest sensitivity, the lowest spatial resolution configuration was used, resulting in typical beam FWHM of $1\secpoint2-1\secpoint6$.  The sensitivity required to detect dust continuum compared to emission lines has been established in previous works at high redshift \citep{Capak2015, Maiolino2015, Inoue2016, Matthee2017, Bethermin2020}. In these studies, it is often found that the spectral scans are slightly more likely to detect an emission line than they are the dust continuum.  The observational strategy of REBELS briefly described here resulted in $3\sigma$ limits in the dust continuum luminosity of $L_{\rm IR}>3\times10^{11}L_{\odot}$ at $z=7$ \citep{Bouwens2021}.  Observations of REBELS targets were obtained from November 2019 to January 2020, with 34 targets having completed their observations, and the remaining targets to be observed in the future. Of these 34 targets, 18 have $>7\sigma$ detections of $[\rm CII]_{158\rm\mu m}$ (described in \citealt{Bouwens2021} and \citealt{Schouws2021prep}) and 13 have a $>3\sigma$ measurement in the dust continuum corresponding to IR luminosities from $L_{\rm IR}=3\times10^{11}L_{\odot}$ to $L_{\rm IR}=1\times10^{12}L_{\odot}$ \citep[described in ][]{Inami2021prep}. Three of the dust-continuum detections are in objects with incomplete spectral scans and thus do not have spectroscopic redshift measurements. The calculation of these IR luminosities is described in Section~\ref{sec:obscuredsfr}.   Observations were reduced and calibrated using the standard ALMA calibration pipeline in \textsc{casa}.  A full description of the observation strategy and data processing techniques is described in \citet{Bouwens2021, Schouws2021prep, Inami2021prep}.

\section{Calculation of the sSFRs}
\label{sec:properties}

In this section we describe the methods used to derive the sSFR for objects in the REBELS sample. This computation includes the estimation of the stellar mass, and derivation of the total SFR.  For the stellar mass we describe several different approaches and describe the systematics included. The total SFR is derived from the sum of both unobscured (UV) and obscured (FIR) components. We describe the methods and uncertainties of both calculations.  Finally, we compute the resulting sSFRs for the REBELS galaxies and compare our derived values to those obtained from SED-fitting of rest-UV and optical photometry. In order to quantify these systematics, we derive galaxy properties using SED models with a variety of assumptions.  Briefly, we test the impact of the assumed dust law in the SED fitting using BEAGLE and comparing the results when Calzetti, SMC, or Milky Way dust is imposed. Additionally, we analyze how the inferred properties derived from SEDs vary for different stellar templates and nebular emission recipes by comparing the output from BEAGLE and Prospector that have identical model setups and constant star-formation histories. Finally, we use Prospector and assume a non-parametric SFH to assess how the assumed SFH impacts the inferred properties. For consistency across all SED models, we adopt log-normal priors for metallicitiy and ionization parameter that are centered at $0.2~\rm Z_{\odot}$ and $\log(U)=-2.5$, with widths of 0.15 and 0.25 dex, respectively, consistent with properties implied by the small sample of rest-UV spectroscopic detections of highly ionized lines at these redshifts \citep[e.g.,][]{Stark2017,Hutchison2019}.

\subsection{Stellar mass}
\label{sec:masses}

%\begin{table*}
%\begin{center}
%\renewcommand{\arraystretch}{1.4}
%\begin{threeparttable}
%\begin{tabular}{rrrrr}
%\toprule
%    Code & SFH   & Dust Law  & Metallicity & Ionization Parameter \\
%          &     &             &     Log-normal Prior       &   Log-normal %Prior     \\
%\midrule

%BEAGLE     & Constant &     Calzetti  &  $\log(Z/Z_{\odot})=-0.7\pm0.15$    & % $\log(U)=-2.5\pm0.25$    \\
%      & Constant &     SMC  &  $\log(Z/Z_{\odot})=-0.7\pm0.15$   &  %$\log(U)=-2.5\pm0.25$     \\
%Prospector     & Constant &     SMC  &  $\log(Z/Z_{\odot})=-0.7\pm0.15$    & % $\log(U)=-2.5\pm0.25$    \\
%      & Non-parametric &     SMC   &  $\log(Z/Z_{\odot})=-0.7\pm0.15$   &  %$\log(U)=-2.5\pm0.25$    \\
%\bottomrule
%\end{tabular}
 
%\end{threeparttable}
% \end{center}
% \caption{Summary of different SED-fitting runs.}
 
%\label{tab:seds}
%\end{table*}

A comprehensive analysis of the methods used to derive stellar masses is presented in \citet{Stefanon2021inprep}, but we provide a brief description here.  Stellar masses were derived using the SED fitting code BayEsian Analysis of GaLaxy sEds  \citep[BEAGLE;][]{Chevallard2016} and Prospector \citep{Johnson2021}. For ease of comparison to previous works, we will adopt the BEAGLE SED models that assume a constant star-formation history (CSFH) as our fiducial set of properties. We also discuss how the adoption of non-parametric star formation histories would influence our conclusions. The BEAGLE tool utilizes the most recent version of the \citet{BC03} stellar population models and  includes a self-consistent treatment of nebular emission based on the photoionization modelling of \citet{Gutkin2016}. These models use a \citet{Chabrier2003} IMF with with stellar masses ranging from  $0.1-300M_{\odot}$. We adopt an SMC dust attenuation law as fiducial but also consider the effects of alternatively assuming a \citet{Calzetti1994} or Milky Way law. 

For each galaxy, the models were fixed at the spectroscopic redshift if available, and otherwise the redshift was allowed to vary. We fit all available photometry from the optical to 
mid-infrared (see \citealt{Bouwens2021} for a full description), and we 
also fit narrowband near-IR photometry where available (e.g.,  \citealt{Endsley2021}). We provide model output values based on the median of the posterior probability distribution, with uncertainties defined as the $16^{\rm th}$ and $84^{\rm th}$ percentiles. Based on this fiducial model setup, we obtain the distribution of stellar masses presented in Figure~\ref{fig:sampleproperties}(c).  This distribution has a median stellar mass of $\log(M_*/M_{\odot})=9.5$ with the full range of stellar masses spanning $\log(M_*/M_{\odot}) = 8.56-10.14$. The median uncertainty on the inferred stellar mass is 0.4 dex.

To explore the impact that different codes and model templates can have, we compare results derived from BEAGLE with those from Prospector with identical initial assumptions. 
For our Prospector fits, we adopt the Flexible Stellar Population Synthesis (FSPS) templates \citep{Conroy2009, Conroy2010} that utilize the MIST isochrones \citep{Choi2016}. We assume a constant star-formation history with a \citet{Chabrier2003} IMF with a high-mass limit of $300M_{\odot}$, and an SMC dust law.   We find broadly consistent results for stellar masses estimated from BEAGLE and Prospector when assuming the same SFH. Specifically, we calculate $\log(M_*/M_{\odot})_{\rm BEAGLE}-\log(M_*/M_{\odot})_{\rm Prospector}$ for each object in our sample, and find a median value of this difference between the two stellar mass estimates of $0.04$ dex. The measured differences scatter about this median with a width of $0.2$ dex, which is within the typical uncertainty determined on the stellar mass. This consistency between the masses inferred using the two codes suggests that the stellar masses are in most cases not strongly sensitive to the assumed model templates (see \citealt{Whitlerinprep} for a more detailed discussion). We find that the choice of the attenuation law also does not strongly impact the derived stellar masses. The median offset between the stellar mass derived assuming the Calzetti and SMC dust laws is just 
0.09 dex, with the SMC law returning modestly smaller masses, on 
average. We find a similar difference when comparing stellar masses inferred assuming SMC and Milky Way dust, with models assuming a Milky Way law yielding stellar masses 0.08 dex larger than those assuming SMC dust on average.  In what follows, we will use the SMC dust law as 
fiducial, but the 
main results would not vary significantly if we had instead 
adopted a Milky Way or \citet{Calzetti1994} law. 

\begin{figure}
    \centering
    \includegraphics[width=1.0\linewidth]{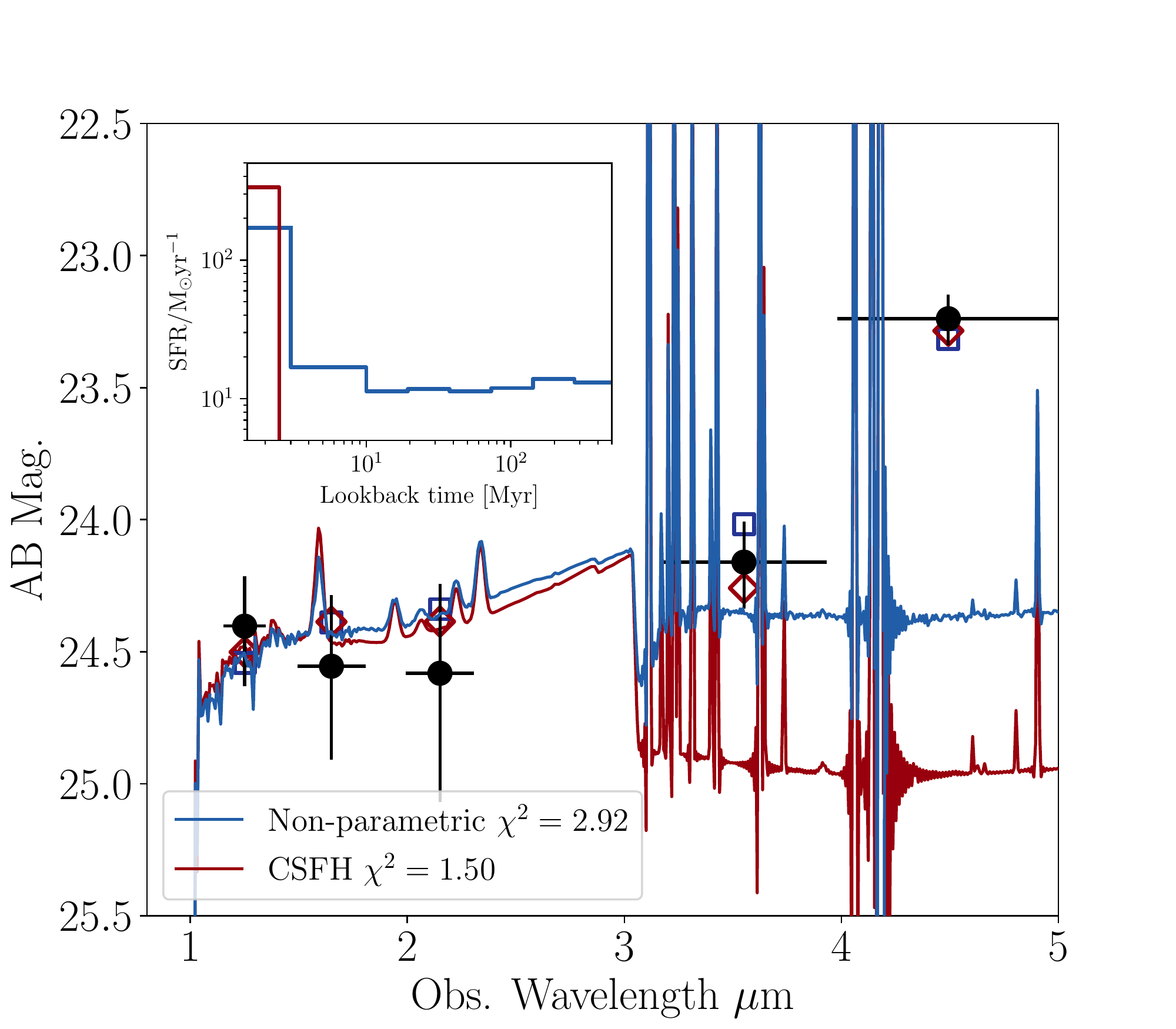}
    \caption{Best-fit SED for REBELS-12 derived using Prospector, and assuming a non-parametric SFH(blue) and a CSFH(red). Observed photometry is shown in black.  Photometric points derived from the best-fit SEDs assuming a non-parametric SFH and CSFH are displayed as blue squared and red diamonds, respectively. This indicates how different assumed SFHs lead to varying estimates of the stellar mass. The inset panel shows the derived SFR for the Prospector non-parametric SFH and CSFH models as a function of lookback time. For this object we find a best-fit stellar mass of $\log(M_*/M_{\odot})=8.93^{+0.55}_{-0.46}$ when assuming a CSFH, and $\log(M_*/M_{\odot})=9.93^{+0.42}_{-0.32}$ for a non-parametric SFH.}
    \label{fig:rebels12}
\end{figure}

The assumed star formation history plays a more significant role 
in the derived mass \citep[e.g.,][]{Lower2020}. Most analyses at very high redshifts have used 
simple parametric star formation histories, such as the CSFH models 
we described previously. It is becoming increasingly clear that 
non-parametric star formation histories can lead to very different solutions \citep[e.g.,][]{Leja2017}. This is particularly important in the reionization era, where a significant fraction of the population appears to be in the midst of a burst (i.e., a recent upturn in star formation; \citealt{Vallini2020}, \citealt{Vallini2021}, \citealt{Legrand2022}, \citealt{Pallottini2022}). This population containing recent bursts faces the classic outshining problem \citep[e.g.,][]{Leja2017, Leja2019}, whereby the light from the recent burst overwhelms that of the older stars which may dominate the stellar mass. Simple parametric models that assume constant star formation return 
very young ages (i.e., $\simeq 3-5$ Myr) for these systems when fitting the rest-UV and 
optical SED \citep[e.g.,][]{Smit2014, Endsley2021} alone. Non-parametric models provide the flexibility to allow star formation at earlier times (i.e., before the burst; Figure~\ref{fig:rebels12}), often leading to significantly higher stellar masses \citep[e.g.][]{Leja2019}. These models thus tend to drive down the sSFRs relative to the parametric CSFH values, with the biggest changes likely to occur in the systems experiencing a recent burst.

To assess the importance of the assumed star 
formation history for our sample, we have fit each of the REBELS galaxies with non-parametric star formation histories using Prospector. The 
approach follows that developed (and described in more detail) in \citet{Whitlerinprep}.  Similar to our approach to the parametric models, we adopt a \citet{Chabrier2003} IMF with an upper mass limit of $300M_{\odot}$ and assume an SMC dust law, with identical priors on ionization parameter and metallicity to those imposed in our fiducial BEAGLE models. % As with the previously described models, the metallicity and ionization parameters were constrained with a log-normal prior centered on $0.2~\rm Z_{\odot}$ and $\log(\rm{U})=-2.5$, with standard deviations of 0.15 dex and 0.25 dex, respectively.
The non-parametric SFHs are composed of eight time bins, with the most recent two bins fixed over the ages of $0-3$ Myr and $3-10$ Myr.  The remaining time bins are distributed logarithmically out to $z=20$. As described in \citet{Whitlerinprep}, the division of the youngest two age bins are required to fit the strongest IRAC excesses seen in the most extreme bursts as is the case in our sample. We additionally adopt the continuity prior built into Prospector that weights against sharp variations in SFR between adjacent time bins (see \citealt{Tacchella2021} for an 
extensive discussion of the influence of different priors in non-parametric models).

\begin{figure}
    \centering
    \includegraphics[width=1.0\linewidth]{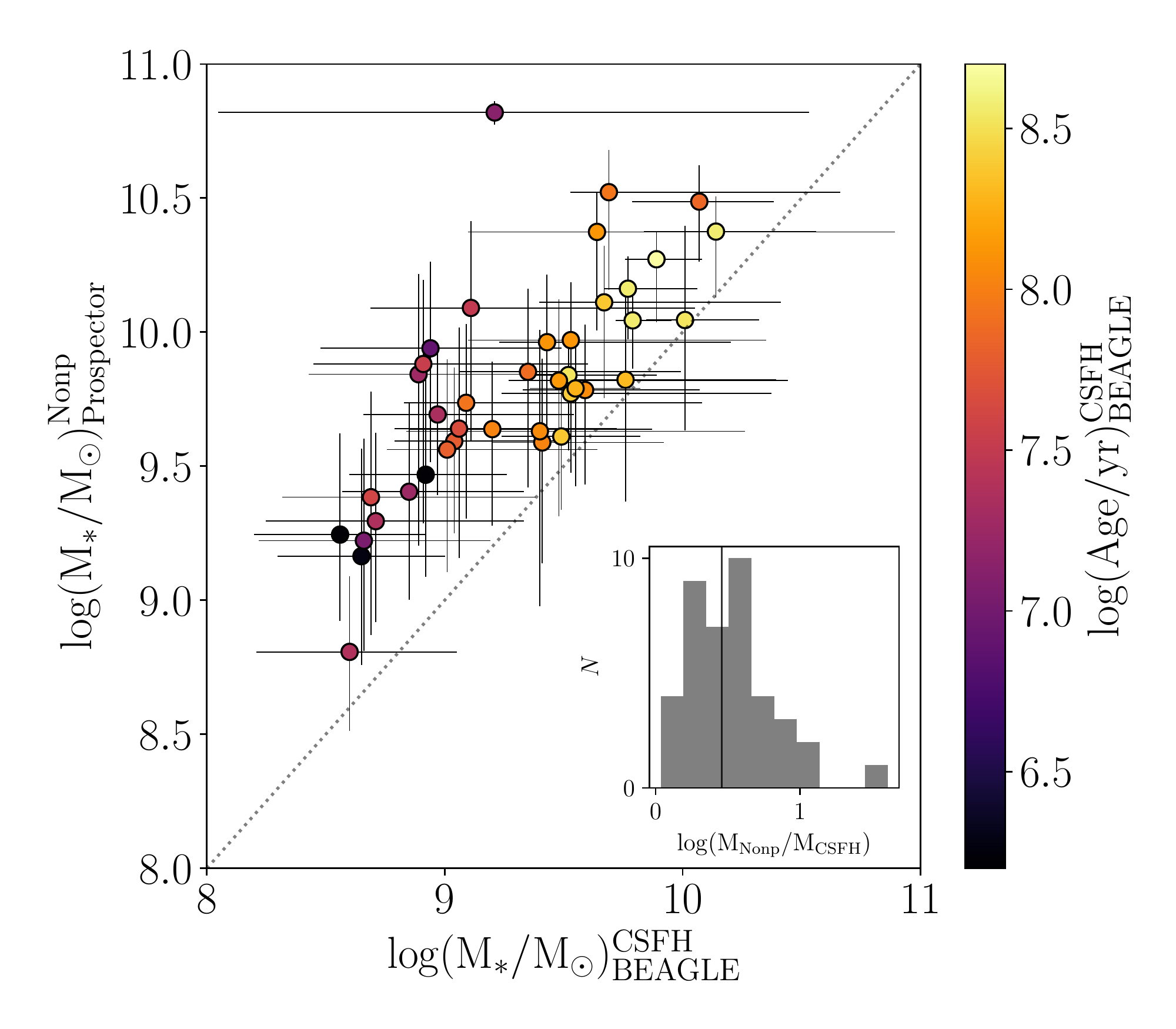}
    \caption{Stellar mass inferred using Prospector and assuming a non-parametric SFH to that inferred using BEAGLE and assuming a CSFH, and color-coded by CSFH age.  The inset panel provides a histogram of the differences in stellar masses derived using these two models. The vertical line within the inset panel indicates the median offset of 0.43 dex.}
    \label{fig:sfh-mass}
\end{figure}

\begin{figure*}
    \centering
    \includegraphics[width=1.0\linewidth]{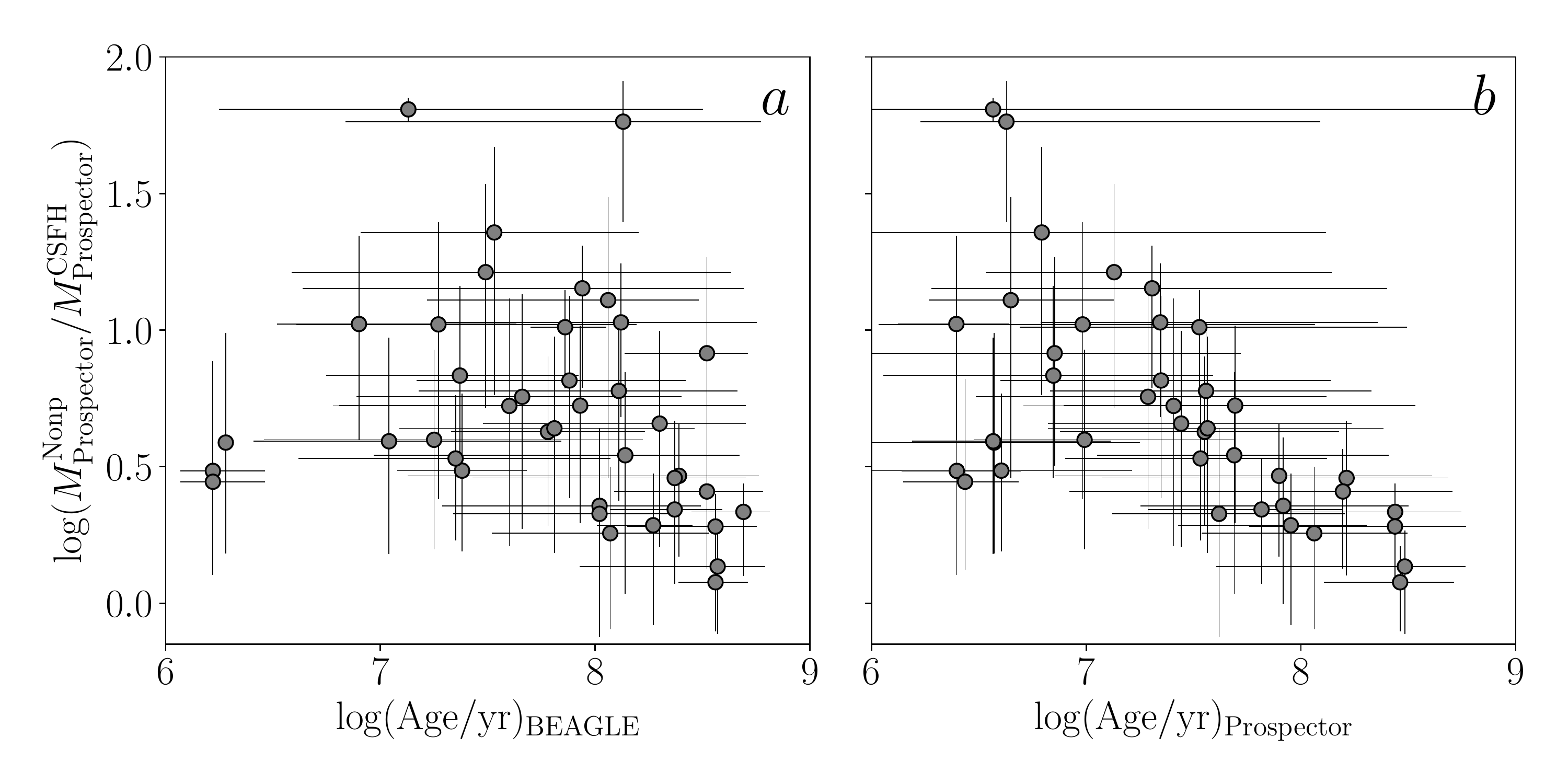}
    \caption{Comparison of stellar mass inferred assuming a non-parametric and constant SFH as a function of galaxy age.  In this figure, age is defined assuming a CSFH using BEAGLE (a), and Prospector (b). In both cases, we find the general trend that the difference between the two stellar mass estimates is greater in galaxies with young ages.}
    \label{fig:mass-comparison}
\end{figure*}

An example of the non-parametric model fits is presented in Figure~\ref{fig:rebels12}. The source shown in this figure, REBELS-12, is among the youngest in the sample, with a best-fit age from BEAGLE CSFH fits of 8 Myr and a stellar mass of $\log(M_*/M_{\odot})=8.93^{+0.55}_{-0.46}$. The non-parametric SFH model
gives a similarly-acceptable fit to the SED, but it suggests a very different past star formation history, with significant low-level star formation at early times and a recent burst. The early star formation in the non-parametric model leads to a stellar mass of 
$\log(M_*/M_{\odot})=9.93^{+0.42}_{-0.32}$, an order of magnitude 
increase over the BEAGLE CSFH value. The same picture holds if we compare to the Prospector parametric CSFH model, in which the non-parametric stellar mass is 10.5 times larger than the parametric CSFH version. One feature that applies throughout the full REBELS sample is that models with non-parametric SFHs are able to supply SEDs that fit the observed data with comparable $\chi^2$ to that of models assuming a CSFH. This illustrates the possibility that these galaxies may be host to more stellar mass than implied by the CSFH models. However, key assumptions in the non-parametric models, such as when the onset of star formation occurred, will require deeper observations to fully constrain. Figure~\ref{fig:sfh-mass} illustrates the difference in inferred stellar mass from the Prospector non-parametric SFH and BEAGLE CSFH models. Across the full sample, we find that the stellar masses inferred from the Prospector non-parametric SFH models are on-average 0.43 dex larger than those derived from the BEAGLE CSFH models. And as we expected, the increase in stellar mass is found to be largest in systems where the CSFH fits lead to low masses and young ages (i.e.,$<\log(M_*/M_{\odot})=9$ and $<$10 Myr, respectively). Figure~\ref{fig:mass-comparison} compares stellar masses inferred using different SFHs as a function of CSFH age for our sample.
For the subset of young objects, the non-parametric models yield an 0.61 dex boost in mass compared to the BEAGLE CSFH models. These changes will clearly affect the sSFRs, particularly for the youngest systems. 
We will come back to discuss the impact of assumed star formation history in Section~\ref{sec:ssfrdist}.

Finally, we consider how the addition of spectroscopic redshifts (a unique aspect of the REBELS sample) improves the reliability of the stellar masses, which is likely to be particularly important at $z\simeq 6.5-7.5$. In this redshift range, emission lines contribute significantly to the IRAC bandpasses, 
and thus the interpretation of the {\it Spitzer}/IRAC fluxes depends sensitively 
on the redshift of the galaxy \citep[e.g.,][]{Labbe2013, Smit2014}. If 
the Spitzer fluxes are interpreted as emission lines, the ages and masses are much 
lower than if the light is produced by stellar continuum. Since the [CII] redshifts in REBELS give the precise position of the nebular lines with respect to the broadband filters, they remove this degeneracy from the fitting process, improving the reliability of the masses. 

\begin{figure}
    \centering
    \includegraphics[width=1.0\linewidth]{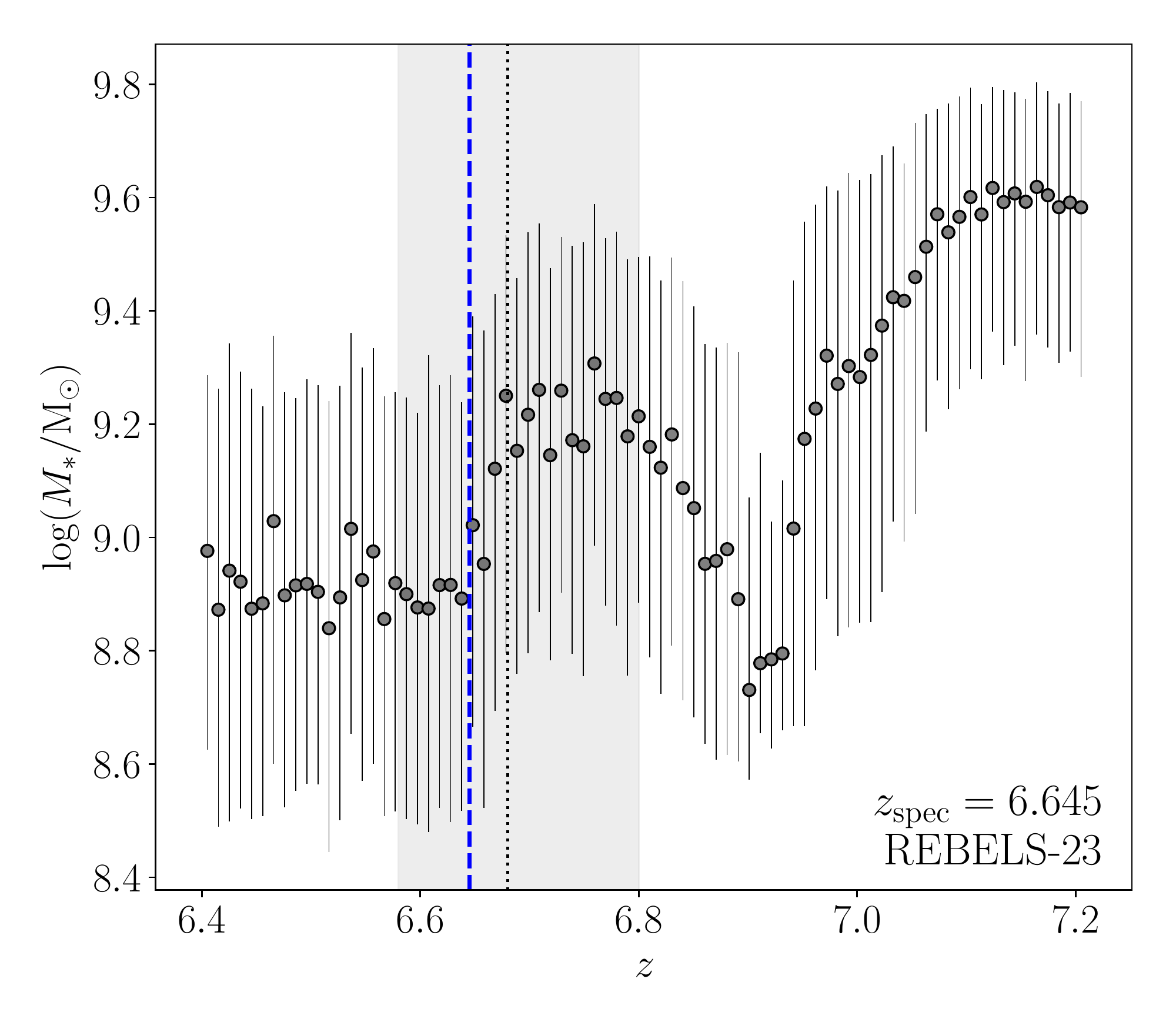}
    \caption{Stellar masses derived from SED fitting as a function of input redshift.  The black vertical dotted line and grey shaded region show, respectively, the best-fit photometric redshift and corresponding uncertainty.  The blue dashed line indicates the spectroscopic redshift measured from $\rm [CII]_{\rm 158\mu \rm m}$.}
    \label{fig:zs}
\end{figure}

To illustrate the magnitude of this effect, we show in Figure~\ref{fig:zs} how the recovered mass changes with redshift for REBELS-23, an object with a [CII] redshift of $z=6.645$ (vertical blue dashed line in the figure) and a reasonably strong (0.6 mag) IRAC excess in [3.6]. The stellar mass we infer when we fix the redshift at its spectroscopic value is 0.4 dex lower than that we infer when we allow the redshift to vary as a free parameter. This change is readily understood looking at the nearly order of magnitude variation in the stellar mass over $6.4<z<7.0$ (Figure ~\ref{fig:zs}) that 
arises as emission lines pass in and out of the IRAC bandpasses. If the photometric redshift is not well constrained, there clearly is potential for substantial error in the stellar mass.

We can quantify the impact of redshift uncertainty in the 23 galaxies in REBELS with spectroscopic redshift determinations, of which 22 have robust [CII] detections \citep[$\rm SNR>5.2$][]{Bouwens2021, Schouws2021prep}, and one object with a low [CII] SNR but has a LyA detection \citep{Endsley2022, Schouws2021prep}.  When we remove the fixed redshift constraint on these objects in the BEAGLE CSFH model fits, we find noticeably larger errors on the recovered stellar masses, with individual systems having uncertainties on the stellar mass that are on average 0.2 dex larger. Additionally, for 5 of the 23 galaxies with spectroscopic redshifts, we find that relaxing the redshift constraint yields an inferred stellar mass that is a factor of 2 discrepant in either direction compared to when the redshift is fixed at the spectrocopic value. In the most extreme case, we find a difference of 1 dex in the mass. However, the average shift across the full sample is only 0.05 dex. Thus while the absence of redshifts in a subset of our sample clearly increases the uncertainty on the derived mass, it is not likely to significantly bias our results.

\subsection{Star-formation Rates}
\label{sec:sfrs}

In this section we describe the methods used to estimate the total SFR for individual galaxies in the REBELS sample. We compute the SFRs by combining inferences of the obscured and unobscured components for each galaxy. As we detail below, unobscured SFRs are calculated using calibrations of SFR/L$_{\rm UV}$ (uncorrected for dust) derived from the SED models presented in \S\ref{sec:masses}, and obscured SFRs are calculated using the ALMA dust-continuum measurements (or upper limits) described in \S\ref{sec:obs}. A complete discussion of the unobscured SFR calculation is presented in \citet{Stefanon2021inprep}.

\subsubsection{The Unobscured SFR}

We first calculate unobscured SFRs for each galaxy using the observed UV continuum luminosity and a conversion factor (SFR/L$_{\rm UV}$) derived from population synthesis models without any dust correction. As galaxies in the REBELS sample span a wide range in ages, to isolate the unobscured SFR we use a SFR/L$_{\rm UV}$ derived individually for each object based on the best-fit CSFH SED model using BEAGLE after the effects of dust have been removed. The age dependence of this calibration is primarily important for young objects, which have a growing B-star population that will not reach an equilibrium for around 100 Myr of constant star formation.  As such, for a fixed $\rm SFR_{\rm UV}$, a younger population will produce a lower UV luminosity compared to an older population (e.g., $\gtrsim 100\rm Myr$) where the massive star population has equilibrated \citep[e.g.,][]{Reddy2012b}. For our sample, 37/40 objects have a $\log(\rm SFR_{\rm UV}/L_{\rm UV}/\rm (M_{\odot}/yr)/(erg/s/Hz))$ in the range $-28.2$ to $-27.9$, with 25 of these 37 objects having the same value to within 0.1 dex.  The remaining 3/40 objects in the sample have a $\log(\rm SFR_{\rm UV}/L_{\rm UV}/(erg/s/Hz)))=-27.4$, due to their young ages. These three systems thus require significantly more unobscured SFR relative to the observed 
UV continuum flux density.

Given the degeneracy between dust and age in the SED fits, the assumed dust attenuation law can also impact the unobscured SFR/L$_{\rm{UV}}$ ratios. However we find that this is not a significant effect for the REBELS sample. The unobscured SFRs derived using models that assume a Calzetti and SMC dust law are very similar, and differ by 0.03 dex on average with corresponding scatter of 0.07 dex. The largest difference in unobscured SFR is 0.4 dex among the full sample. Similarly, unobscured SFRs inferred assuming Milky Way dust are 0.04 dex larger than those from models with an SMC law.

Finally, we also must consider whether the non-parametric SFHs 
influence the unobscured SFR/L$_{\rm{UV}}$ ratios that the 
Prospector models return. When considering unobscured SFRs derived from non-parametric SFH models, we adopt a value for the SFR that is averaged over the past 10 Myr of the SFH.  For the vast majority of our sample (34/40), the average offset between the Prospector CSFH and non-parametric unobscured SFR measures is minimal, such that they agree within the uncertainties with an overall average difference of 0.02 dex.  The remaining subset have larger SFRs derived when assuming a CSFH, with the largest offset being a factor of $\times 7$ difference. However, the agreement on average for the sample indicates that both SFHs typically provide broadly consistent measures of the unobscured SFR.

\subsubsection{The Obscured SFR}
\label{sec:obscuredsfr}

The obscured component of the star-formation rate is inferred from the ALMA-based constraints on the IR continuum luminosities integrated over $8-1000\mu m$, $L_{\rm IR}$. A detailed description of this obscured SFR derivation is provided in \citep{Inami2021prep}, however we present a brief summary here. We then discuss our method for constraining obscured SFR in those sources lacking detections in the IR 
continuum.

For the 16 sources in REBELS with dust continuum detections, we scale the dust continuum luminosity at rest-frame $158~\mu m$ or $88~\mu m$
to the total IR luminosity by assuming a modified black body with $\beta_d=2.0$ and a dust temperature of T$_{\rm{d}}=47$ K, obtained assuming Milky Way-like dust, which has been shown to reproduce the IR properties of REBELS objects \citep{Sommovigo2022, Inami2021prep, Ferrara2022}. For comparison, this dust temperature is slightly higher than what is found for ALPINE at 
$z\simeq 5-6$ \citep[$\rm T_{\rm d}=43$ K;][]{Bethermin2020}. We adopt this  temperature based on analysis of the 13 galaxies in REBELS with [CII] and dust-continuum measurements for which dust temperatures can be constrained using the method described in \citet{Sommovigo2022}. The objects span a range of temperatures from $39-58$ K with the median value of 47 K.  
This chosen temperature results in a scaling of $L_{\rm IR}\equiv 14^{+8}_{-5}\nu L_{\nu}$ ($L_{\rm IR}\equiv 8^{+1}_{-4}\nu L_{\nu}$), where $\nu$ is the frequency corresponding to the [CII] $158\mu m$ ([OIII]$88~\mu m$) line. The uncertainty on this conversion factor reflects the variation in dust temperatures established for this subset of REBELS sources \citep{Sommovigo2022}. We note that the median temperature is within the range of dust temperatures measured for galaxies at similar redshifts \citep{Knudsen2017, Bowler2018, Hashimoto2019, Bakx2021}. The increased stellar masses inferred from non-parametric SFH models impact the derivation of this conversion factor, yielding a value of $L_{\rm IR}\equiv 12^{+4}_{-2}\nu L_{\nu}$. While this different conversion factor results in slightly lower IR luminosities, we use this calibration when calculating IR luminosities (and therefore obscured SFRs) in the context of non-parametric SFH models.

The obscured SFRs are then calculated from this quantity using the conversion $\rm SFR_{\rm IR}=1.2\times 10^{-10}~L_{\rm IR}/L_{\odot}$ obtained from \citet{Madau2014}, where here we have assumed a CSFH age of 100 Myr, corresponding to the average for the REBELS sample. As noted in \S3.2.1, there are three galaxies in the REBELS sample with very young ages derived from UV and optical photometry. Since it is not clear that these young ages are also associated with the component of the galaxies dominating the FIR, we do not alter the conversion factor for these three systems. Doing so 
would modestly increase the obscured SFR in these systems but would not significantly 
impact the overall results of the full sample. The methodology of computing total IR luminosities and SFR$_{\rm IR}$ described above is comparable to that taken in other analyses of galaxies at high-redshift and theoretical models \citep[e.g.,][]{Bethermin2020, Sommovigo2021}.

For the REBELS objects with continuum detections, this procedure results in measured total IR luminosities that span $2.8 - 15\times 10^{11}~\rm{L}_{\odot}$. Based on our assumed calibration, these IR luminosities yield obscured SFRs ranging from $34~\rm M_{\odot}\rm~yr^{-1}$ to $180~\rm{M}_{\odot} ~\rm yr^{-1}$. For these objects, the obscured fraction is typically high, with $\rm SFR_{\rm IR}/\rm SFR_{\rm tot}$ ranging from $0.58$ up to $0.92$, with a median obscured fraction of $0.72$, consistent with those found in \citet{Stefanon2021}. An additional complication is that while measurements in the FIR dust continuum provide a direct probe of the obscured star formation, the translation between these two quantities is potentially subject to uncertainties. For example, assuming a dust temperature that is 10 K lower than our assumed 
value (47 K) would affect the scaling between L$_{\rm IR}$ and L$_{\nu}$, resulting in lower estimates of $\rm SFR_{\rm IR}$ by $0.3$ dex \citep[e.g.;][]{Bowler2018}. Based on the temperature distributions independently derived in \citet{Sommovigo2022} and \citet{Ferrara2022}, which are consistent with our chosen median value of 47 K, it is unlikely that the entire sample has such low dust temperatures.  Additionally, such low temperatures would increase the tension with measurements of dust production at $z\sim7$ \citep[e.g.,][]{Sommovigo2020, Dayal2022}. Nonetheless, deviations from this median dust temperature in individual systems can potentially lead to some scatter around the true obscured SFRs.

Finally we discuss our procedure for constraining the level 
of obscured SFR in the 24 sources in REBELS that  
do not have a dust continuum detections. For these sources, the upper limits on the dust continuum can be translated into 
an upper limit on the obscured star formation rate. To 
obtain these constraints, we first median combine the non-detections, 
splitting the sample into two equal bins based on their UV slope. We choose UV slope bins because of the relation between UV-slope and $L_{\rm IR}/L_{\rm UV}$ \citep[IRX; e.g.,][]{Meurer1999, Casey2014}, which has been evaluated in high-redshift samples \citep[e.g.,][]{Capak2015, Bouwens2016, Reddy2018, Fudamoto2020} and will be presented in \citet{Bowlerinprep} for the REBELS sample. We split the bins by the sample median value of $\beta=-2.04$.  The bluer bin comprising 13 galaxies with a median $\beta=-2.2$, and the redder bin with a median $\beta=-1.7$ containing 11 galaxies. This stacking procedure potentially introduces some bias such that the objects with redder UV slopes contribute more to the stacked FIR luminosity. Additionally, uncertainties in UV-slope measurements may result in significant scatter between these two bins \citep{Bowlerinprep}. However, this presents an improvement over stacking the full sample of non-detections.  For the 11 galaxies in the redder bin, we measure a peak flux of $24\pm6\rm \mu Jy$. For the stack of 13 bluer galaxies, we find no detection and measure a $3\sigma$ upper limit of $14\rm \mu Jy$. We convert these flux constraints to an average $L_{\rm IR}$ (or upper limit in the case of the bluer bin) using the previously described conversion factor and including the corresponding uncertainty, and then calculate an average IRX. For the redder bin, we achieve an average $\log(\rm IRX)=-0.02$, and for the bluer bin we obtain an upper limit of $\log(\rm IRX) < -0.15$. For each of the undetected sources, we then calculate $L_{\rm IR}$ from the average IRX resulting from the stacks, and then derive an obscured SFR (or limit) using the method described above.

\subsection{Synthesis of sSFRs}
\label{sec:ssfrs}

\begin{figure*}
    \centering
    \includegraphics[width=1.0\linewidth]{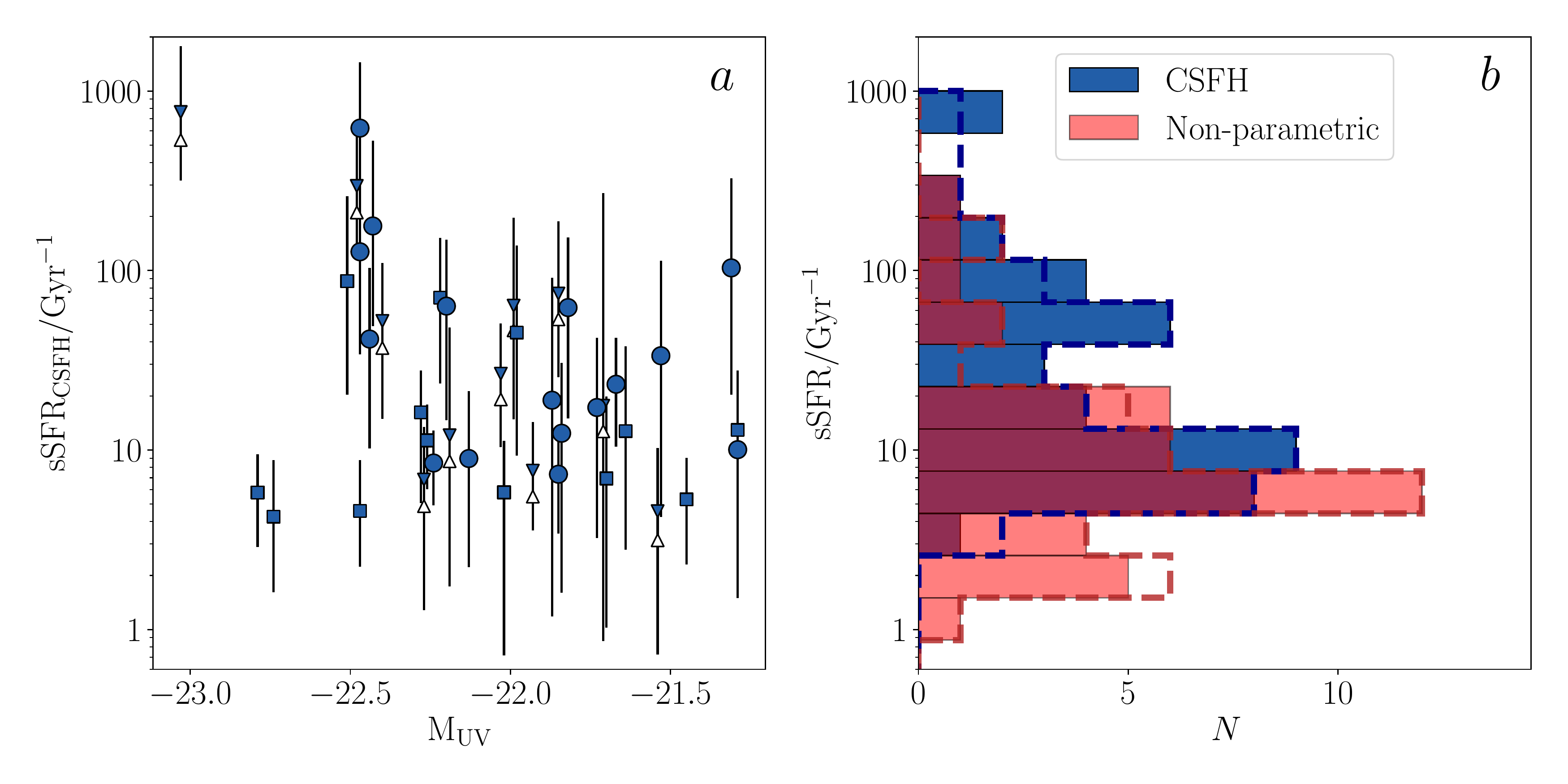}
    \caption{Specific star-formation rates of the REBELS sample {\it a:} sSFR$_{\rm CSFH}$ as a function of absolute UV magnitude.  The large points represent objects in our sample for which the obscured SFR has been measured from the FIR dust continuum.  The squares show values calculated based on a stacked detection of the dust continuum. For the sample of galaxies without dust-continuum detections and blue UV-slopes, we show the two limiting cases for the sSFR, where the amount of obscured star formation is set to the upper limit (blue triangles), and those where the obscured star formation is set to zero (white triangles).  {\it b: }Distribution of specific SFR for the REBELS sample calculated assuming a constant SFH (blue) and non-parametric SFH (red). We provide the distributions described above, where obscured star formation is set to their upper limit (filled histogram), and the case where there is assumed to be no obscured star formation in objects without dust-continuum detections and blue UV-slopes (dashed line). For the CSFH-derived (non-parametric) values, these distributions are characterized by a median of $\rm sSFR_{\rm{CSFH}} =18_{-5}^{+7}~\rm Gyr^{-1}$ ($\rm sSFR_{\rm{Nonp}}=7.1_{-2.2}^{+2.8}~\rm Gyr^{-1}$), and $\rm sSFR_{\rm{CSFH}} =16_{-5}^{+7}~\rm Gyr^{-1}$ ($\rm sSFR_{\rm{Nonp}}=6.2_{-1.8}^{+2.4}~\rm Gyr^{-1}$), respectively.}
    \label{fig:ssfrdistribution}
\end{figure*}

In the previous sections we described the derivation of stellar mass and SFR for the individual galaxies in the REBELS sample.  Here we combine these quantities to compute sSFRs and discuss systematics that may affect the overall sSFR distribution.  

For the 16 objects in REBELS that have individual dust continuum measurements, we measure a median $\rm sSFR_{\rm{CSFH}}$ of $27^{+24}_{-11}~\rm Gyr^{-1}$. The requirement of a dust continuum detection may preferentially select objects that are most intensely forming stars. To understand this effect on our sSFR distribution, we  examine the 
24 systems lacking individual FIR continuum detections. The obscured SFRs for this subset  are derived based on stacked measurements of their dust continuum with the 
sample split into two bins of UV slope. As previously described, the bluer of the 
two bins (centered at $\beta=-2.2$) is not detected in the continuum stack. We bracket the sSFRs of these 13 galaxies considering two limiting cases. The upper bound comes from setting the 
obscured SFRs of this subset to the $3\sigma$  upper limit implied by the stack ($\log(\rm IRX)<-0.15$), and the lower bound comes from 
setting L$_{\rm{IR}}$=0.  With this approach we derive the sSFR of the 24 galaxies in REBELS that are undetected in 
the dust continuum. The median sSFR of this subset is 
between $\rm sSFR_{\rm{CSFH}}=11~\rm Gyr^{-1}$ 
and $\rm sSFR_{\rm{CSFH}}=13~\rm Gyr^{-1}$, with the range set by the two bounds discussed above. As expected, these numbers indicate that the subset of REBELS 
sources lacking detection in the dust continuum have slightly 
lower sSFR than those with FIR detections.

We can now quantify the sSFR distribution of the entire 
40 galaxies in the REBELS sample. The individual sSFR 
values for our fiducial CSFH models are shown in Figure~\ref{fig:ssfrdistribution}.
To calculate the median of the distribution, we again consider two limiting cases for the subset of 13 galaxies described above. This procedure suggests the median of the full sample 
ranges between $\rm sSFR_{\rm{CSFH}} =16_{-5}^{+7}~\rm Gyr^{-1}$ and $\rm sSFR_{\rm{CSFH}} =18_{-5}^{+7}~\rm Gyr^{-1}$. These values are derived using a bootstrap Monte-Carlo method, where we randomly select 40 objects with replacement from the REBELS sample, perturb their stellar masses, unobscured SFR, and obscured SFR by the associated uncertainties for each source, and calculate the median.  This process is repeated 1000 times, and the uncertainty is defined at the $16$th and $84$th percentile of the resulting distribution of median sSFRs.

The sSFR values quoted above are valid for the assumed constant star formation history. This is consistent with what has typically been used in the literature at high redshift and thus serves as our best benchmark for investigating the evolution of sSFR. However as we showed in \S3.1, non-parametric star-formation histories can significantly 
alter the sSFRs. The differences arise primarily due to 
changes in the stellar masses (see Figure~\ref{fig:mass-comparison}), as the 
average SFRs vary much less significantly (see \S3.2). 
For simplicity, we thus calculate non-parametric sSFRs 
for the REBELS sample by combining the total UV+IR star formation rates (see \S3.2) with the non-parametric stellar masses (see \S3.1).

As expected from our discussion in \S3.1, the changes when non-parametric models are invoked are most significant for the lowest mass (and youngest) sources in the sample (see Figure~\ref{fig:mass-comparison}). Considering the entire REBELS sample, the median sSFR inferred using non-parametric SFHs ranges between $6.2_{-1.8}^{+2.4}~\rm Gyr^{-1}$ and $7.1_{-2.2}^{+2.8}~\rm Gyr^{-1}$, where this range is determined using the same assumptions on the non-detections described above. These values are, respectively,  
0.38 and 0.36 dex lower than the CSFH values derived using our fiducial assumptions. We will discuss how the lower sSFRs implied by non-parametric models may impact our conclusions in the following sections.

\subsection{Comparison of UV+IR and SED-based sSFRs}

The majority of sSFR determinations at $z>7$ have been derived from SED fitting of UV and optical photometry. In the next several years, {\it JWST} will deliver many more UV+optical sSFRs in 
this redshift range. The REBELS sample allows us to investigate how these UV+optical SED-based determinations compare to those derived when FIR constraints are available.
For each source in REBELS, we measure the SED-based UV+optical sSFR using our fiducial BEAGLE models and compare to the UV+IR measurements. We find that SFRs inferred directly from the UV+IR are elevated relative to estimates from the UV+optical SED. 
This in turn leads to larger sSFR values when the dust continuum constraints are 
utilized.  In particular, we find that the median sSFR based on BEAGLE UV+optical SED fits for the REBELS sample is $\rm sSFR=9.5^{+2.4}_{-2.0}~\rm Gyr^{-1}$, which is $0.28$ dex lower than the values we derive in \S3.3 making use of the FIR continuum 
constraints. The SED-based median sSFR decreases to $\rm sSFR=8.5^{+2.2}_{-1.8}~\rm Gyr^{-1}$ when
a Calzetti law is adopted instead of SMC. This implies a significant offset between the sSFR we 
derive from the traditional UV+optical SED fitting techniques and what we 
derive when the dust continuum is available. We note that this offset is not sensitive 
to the form of the star formation history, as we find similar results using the non-parametric models. The assumed dust temperature does play a 
role. As discussed in \S3.2.2, lower dust temperatures would bring down the obscured star 
formation rates. However for the two estimates to match, we would require an average dust temperature below 40 K (see \S3.2.2), lower than the range predicted for the REBELS sample \citep{Sommovigo2022}. 
Future observations are required to confirm and investigate this offset.  We will discuss possible physical effects that may contribute in \S\ref{sec:implications}.

%=============================================
%
%         RESULTS
%
%==============================================
\section{Results}

In this section, we use the UV+IR based SFRs and stellar masses to constrain the $z\simeq 7$ star forming main sequence and the distribution of sSFRs in the REBELS sample. We close by exploring the relationship between the sSFR and the UV luminosity as well as [OIII]+H$\beta$ EW.

\subsection{Star forming main sequence at $z\sim7$}
\label{sec:mainsequence}

\begin{figure*}
    \centering
    \includegraphics[width=1.0\linewidth]{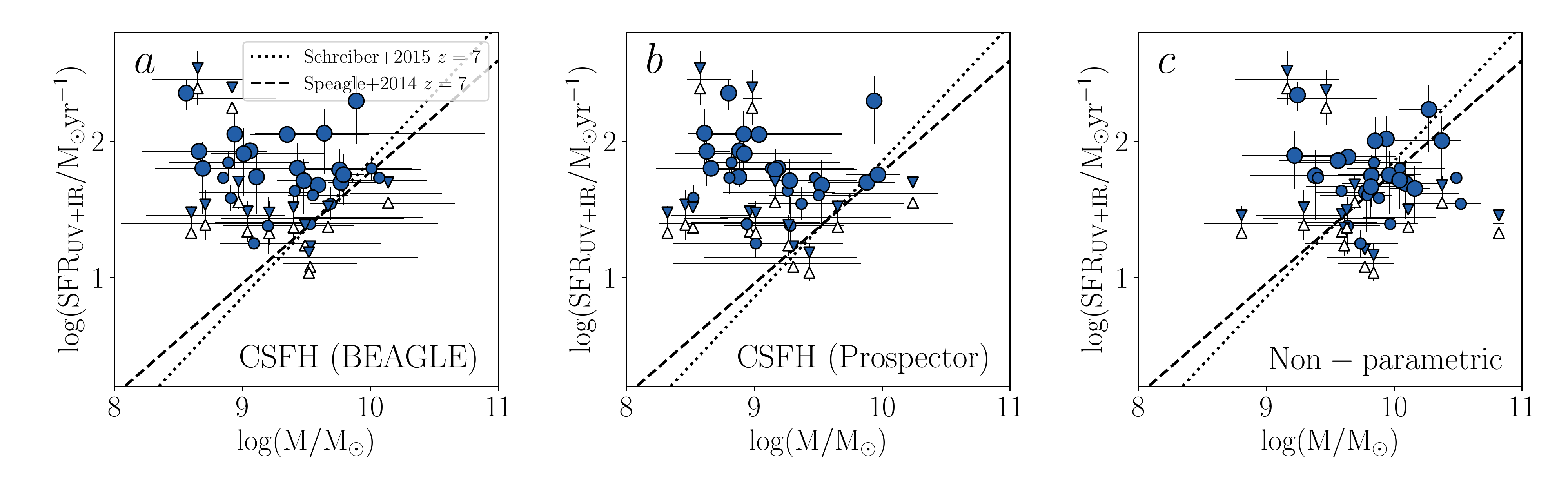}
    \caption{SFR vs. stellar mass for REBELS galaxies at $z\sim7$. The symbols display measurements using the same scheme as in Figure~\ref{fig:ssfrdistribution}(a). Each panel displays the main sequence extrapolated to $z=7$ from \citet{Speagle2014} and \citet{Schreiber2015}. {\it a}: SFR vs. stellar mass derived using stellar masses inferred from BEAGLE and assuming a CSFH. {\it b: }SFR vs. stellar mass derived using stellar masses inferred from Prospector and assuming a CSFH. {\it c: }SFR vs. stellar mass derived using stellar masses inferred from Prospector and assuming a non-parametric SFH. }
    \label{fig:ms}
\end{figure*}

In Figure~\ref{fig:ms}, we present the REBELS star forming main sequence derived using stellar masses inferred from three different SED-fitting prescriptions described above (BEAGLE CSFH, Prospector CSFH, Prospector non-parametric). The SFRs are calculated from the UV+IR measurements that we described in \S\ref{sec:sfrs}. 
We compare the REBELS galaxies to the star forming main sequences presented in \citet{Speagle2014} and \citet{Schreiber2015}. In each 
case, we extrapolate their relations to $z=7$. Crucially, these two references utilized direct constraints on the obscured SFR from measurements in the FIR, providing an appropriate comparison to our sample.
When we assume a constant star-formation history, 
the low mass REBELS galaxies are well above the predicted 
main sequence, with nearly the same SFRs as those 
in the sample with larger masses. 
Specifically, galaxies with CSFH masses inferred using BEAGLE that are $\log(\rm M_*/M_{\odot}) < 9$ have, on average, SFRs that are elevated above the main sequence defined by \citet{Speagle2014} by a factor of 11, and that of \citet{Schreiber2015} by a factor of 14. At stellar masses $\log(M_*/M_{\odot})_{\rm BEAGLE}>9.5$, the galaxies show better consistency with the comparison main sequences, however some objects still have SFRs that lie above by up to 0.5 dex. At the median stellar mass of our sample ($\log(\rm M_*/M_{\odot})=9.5$) we establish an average SFR of $47~\rm M_{\odot}/yr$.  We find nearly identical results for the main sequence derived using CSFH stellar masses from Prospector. As with BEAGLE, galaxies at $\log(\rm M_*/M_{\odot}) < 9$ are elevated above the main sequence of \citet{Speagle2014} and \citet{Schreiber2015} by factors of 11 and 16, respectively.

The star forming main sequence derived when a non-parametric SFH is assumed looks distinctly different to that described above (see Figure~\ref{fig:ms}). As described in \S\ref{sec:masses}, models that assume a non-parametric SFH allow for the inclusion of older stellar components in cases where the light is dominated by a recent burst. The result is an overall increase in stellar mass compared to the CSFH models, which more strongly affects objects at the young and low-mass end of the CSFH distribution (Figure~\ref{fig:mass-comparison}).  With the 
non-parametric masses, we find improved consistency between the REBELS galaxies and the extrapolated $z=7$ main sequences of \citet{Speagle2014} and \citet{Schreiber2015}.  

The limited dynamic range in the stellar mass  makes it challenging to derive precise fitting functions for the 
star forming main sequence in the REBELS sample. In particular, 
it is difficult to establish the slope of the main sequence at $z\sim7$ with only REBELS galaxies.  However, we can estimate the normalization of the REBELS main sequence by fixing the slope to that determined by \citet{Speagle2014} and \citet{Schreiber2015} at $z=7$ of $\log(\rm SFR)/\log(\rm M)=0.82$ and $1.0$, respectively.  Using this method, for masses derived assuming a CSFH, we find a main sequence normalization that is 0.46 and 0.50 dex higher SFR at fixed stellar mass compared to \citet{Speagle2014} and \citet{Schreiber2015}, respectively.  In contrast, we find much better agreement when comparing to our non-parametric stellar masses. Using these stellar masses, we find normalization offsets are only 0.06 and 0.02 dex higher in SFR at fixed stellar mass, corresponding to main sequences of $\log(\rm SFR/M_{\odot}yr^{-1}) = 0.82\times\log(\rm M_*/M_{\odot})-6.36$ and $\log(\rm SFR/M_{\odot}yr^{-1}) = \log(\rm M_*/M_{\odot})-8.12$. 

\subsection{The sSFR distribution}
\label{sec:ssfrdist}

Here we consider the range of sSFRs in the REBELS sample.
Figure~\ref{fig:ssfrdistribution}(b) shows the distribution of sSFRs obtained using the fiducial BEAGLE CSFH models.  As described in \S3, we derived obscured SFRs for objects without individual dust-continuum detections through a stacking analysis of the IR continuum in two bins separated by UV continuum slope. The bluest bin did not yield a detection in the stack, so we considered two limiting cases that 
bracket the range of obscured SFR in these systems (see \S3.2.2 for more information). The two corresponding sSFR distributions are shown in Figure~\ref{fig:ssfrdistribution}(b) in blue and as a black dashed line respectively.  As described in \S\ref{sec:ssfrs}, the two distributions have similar medians of $\rm sSFR_{\rm{CSFH}} =18_{-5}^{+7}~\rm Gyr^{-1}$ for the upper limiting case and $\rm sSFR_{\rm{CSFH}} =16_{-5}^{+7}~\rm Gyr^{-1}$ for the lower limiting case. The adoption of non-parametric 
SFHs increases the stellar masses (mostly at the low mass 
end), which in turn reduces the sSFRs. For our non-parametric masses, we similarly determine the sSFR distribution for the two scenarios describing objects in the bluest FIR stack, and find a median  $\rm sSFR_{\rm Nonp} =7.1_{-2.2}^{+2.8}~\rm Gyr^{-1}$ for the upper limiting case, and $\rm sSFR_{\rm Nonp} =6.2_{-1.8}^{+2.4}~\rm Gyr^{-1}$  for the lower limiting case. A summary of median sSFRs derived for the several samples and assumptions is provided in Table~\ref{tab:ssfrs}.

We additionally consider the scatter in the sSFR distribution, which is sensitive to  variations in the star formation histories of galaxies at a fixed mass. For our fiducial CSFH models, we measure a scatter, defined as the bi-weight scale of the distribution, of 0.49 dex for both of the limiting cases considered for the IR  non-detections. We note that the posterior on the sSFR in individual REBELS systems implies uncertainties that are comparable to the scatter quoted above. As such, we cannot robustly estimate the intrinsic scatter of sSFRs for this sample. We also consider the scatter in sSFR 
for the non-parametric SFHs.  As the changes in the sSFR 
distribution are typically more significant for galaxies with high $\rm sSFR_{\rm CSFH}$ (i.e., the young and low mass 
galaxies in the CSFH modeling), we expect the use of non-parametric models to also affect the width of the resulting sSFR distribution. Indeed this is the case for our sample. We find that adopting non-parametric SFHs results in a scatter of 0.37 dex, reducing the width of the sSFR distribution relative to CSFH models by 0.12 dex. 

%Using stellar masses derived assuming a non-parametric SFH has the additional effect of reducing the range of stellar mass spanned by the sample which, in turn, reduces the width of the sSFR distribution.  

\begin{table*}
\begin{center}
\renewcommand{\arraystretch}{1.4}
\begin{threeparttable}
\begin{tabular}{cccccc}
\toprule
    SED model & Stellar mass range & $\rm M_{\rm UV}$ range &  SFR method & $L_{\rm IR}$ sample & Median sSFR    \\
                &            &           &             & & $[\rm Gyr^{-1}]$ \\
\toprule

BEAGLE CSFH     & All & All  &            UV+IR  &      All      & $18^{+7}_{-5}$\ ($16^{+7}_{-5}$)        \\
                & All & All  &           SED     &      All      & $9.5^{+2.4}_{-2.0}$               \\
                & All & All  &           UV+IR   & Detections only & $27^{+24}_{-11}$              \\
                & All & All  &           UV+IR   & Non-detections only & $13^{+7}_{-5}$ ($11^{+6}_{-4}$)             \\
                & All & $\rm M_{\rm UV} < -22.0$  &  UV+IR &All & $33^{+70}_{-23}$         \\
                & All & $\rm M_{\rm UV} \ge -22.0$  &  UV+IR &All & $15^{+22}_{-9}$         \\
                & $9.6 < \log(\rm M_*/M_{\odot}) < 9.8$ & All & UV+IR &All & $8.0^{+3.0}_{-2.3}$        \\
\midrule[0.1pt]
Prospector Non-parametric     & All& All   &UV+IR &All & $7.1^{+2.8}_{-2.2}$ ($6.2^{+2.4}_{-1.8}$)        \\
                & $9.6 < \log(\rm M_*/M_{\odot}) < 9.8$ & All & UV+IR &All & $6.2^{+2.4}_{-1.8}$        \\

 \bottomrule
 \end{tabular}
 
\end{threeparttable}
 \end{center}
 \caption{Summary of sample median sSFR determinations. Median sSFRs derived assuming no obscured star formation in objects without dust-continuum measurements are given in parentheses.}
 
\label{tab:ssfrs}
\end{table*}

\subsection{Dependence of sSFR on M$_{\rm{UV}}$ and [OIII]+H$\beta$ EW}
The REBELS sample allows us to investigate how the sSFR at $z\simeq 7$ depends on various galaxy properties and observables. Here we consider whether there are any trends 
between sSFR and the absolute UV magnitude and the [OIII]+H$\beta$ EW. We first consider the relationship between sSFR and M$_{\rm UV}$.
Figure~\ref{fig:ssfrdistribution}(a) shows the sSFR as a function of absolute UV magnitude over the range spanned by REBELS galaxies of $-23.0 \le \rm M_{\rm UV} \le -21.3$. The uncertainties in the individual sSFR measurements includes errors in the obscured SFR and unobscured SFR but are dominated by uncertainties in the stellar mass. Objects without individual detections in the dust continuum are shown as smaller points. As described above, we are not able to directly measure the obscured SFR for a subset of our sample lacking individual and stacked dust continuum detections.  We therefore provide sSFRs in the limiting cases, where we set the obscured SFR to its upper limit (blue triangles) and lower limit within these objects (white triangles). We note that the objects without individual dust-continuum measurements span roughly the same range of M$_{\rm{UV}}$ as the objects with detections.

We calculate the median sSFR of REBELS galaxies in two bins of M$_{\rm UV}$ delineated at the median value of the sample. This calculation yields a median  $\rm sSFR_{\rm{CSFH}} =33_{-23}^{+70}~\rm Gyr^{-1}$  for the bin centered at $\rm M_{\rm UV}=-22.4$, and a median  $\rm sSFR_{\rm{CSFH}} =15_{-9}^{+22}~\rm Gyr^{-1}$  for the bin at $\rm M_{\rm UV}=-21.7$.  Furthermore, a Spearman correlation test results in a correlation coefficient of $r_s=-0.22$ and a p-value of $0.17$, consistent with no correlation. Additionally, when considering the sSFRs derived assuming non-parametric SFHs (as described in \S3.3), we 
also do not find evidence for a significant relationship between sSFR and $\rm M_{\rm UV}$. Based on these tests, and possibly due to the large uncertainties present for the individual sSFR measurements, we do not observe any significant correlation with sSFR and M$_{\rm UV}$.  However, as the REBELS sample does not span a large dynamic range in M$_{\rm UV}$, and is composed of the most UV-bright galaxies, this result does not preclude such a correlation toward lower UV luminosities.

\begin{figure}
    \centering
    \includegraphics[width=1.0\linewidth]{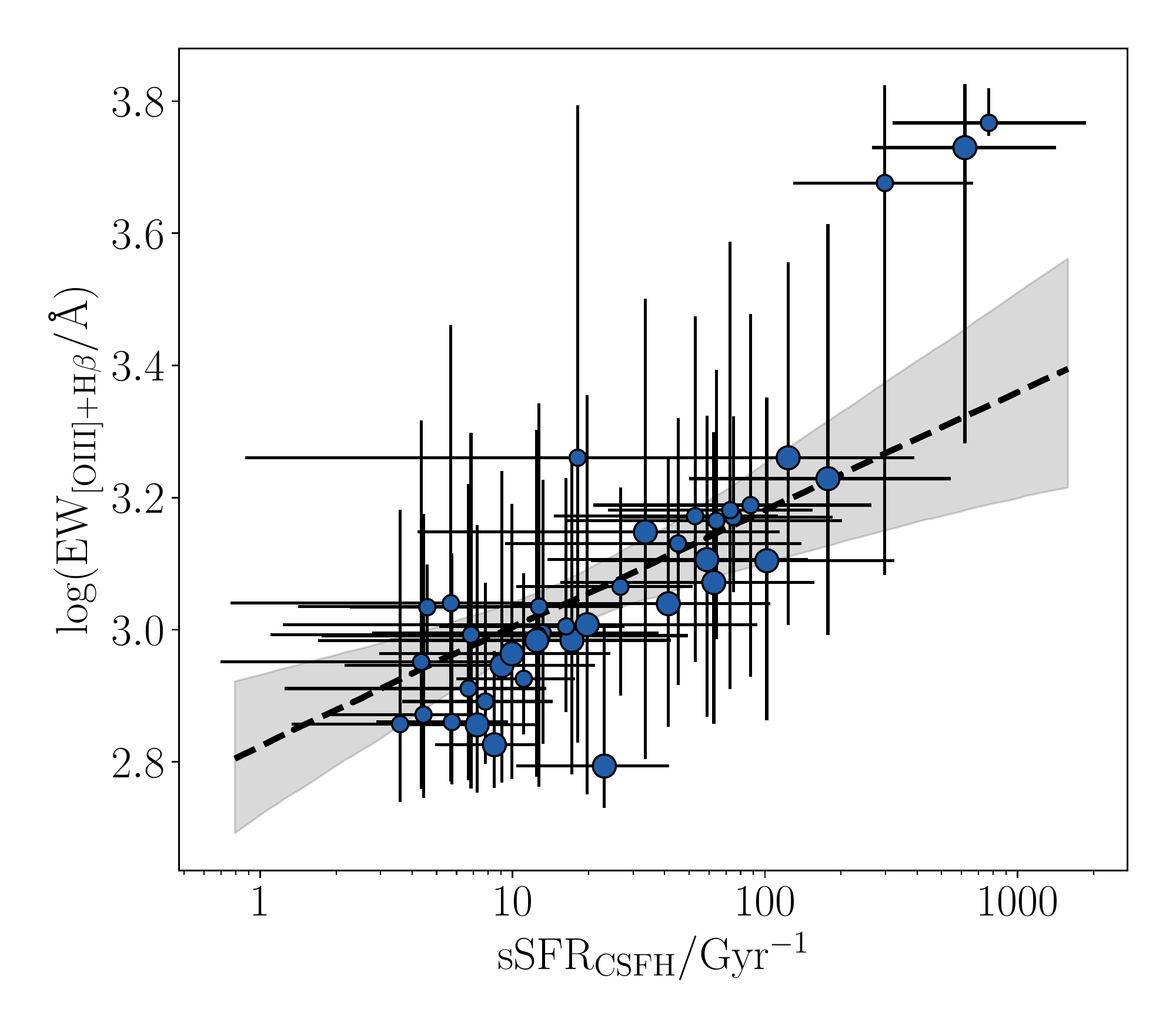}
    \caption{[OIII]+H$\beta$ equivalent width derived from our fiducial BEAGLE SED models as a function of sSFR$_{\rm CSFH}$.  We observe a clear correlation between these two parameters such that the objects with the highest EWs also have the highest sSFRs. }
    \label{fig:ssfr-vs-ew}
\end{figure}

The [OIII]+H$\beta$ EW has been derived from {\it Spitzer}/IRAC flux excesses in many reionization-era galaxies. For REBELS systems, we constrain the line properties through our SED modeling with BEAGLE \citep{Bouwens2021, Stefanon2021inprep}. 
As the [OIII]+H$\beta$ EW is the ratio of nebular emission 
line luminosities (powered by O stars) and the rest-optical 
continuum (sensitive to presence of A stars), we expect 
it to correlate with the sSFR when observed over a large enough dynamic range. 
We investigate the relationship between these two quantities for galaxies in the REBELS sample. The [OIII]+H$\beta$ EW derived assuming a CSFH with BEAGLE is shown as a function of sSFR for objects in the REBELS sample in Figure~\ref{fig:ssfr-vs-ew}. There is a clear correlation between the two parameters, similar to that seen at low-redshift  \citep[e.g.,][]{Amorin2015} and at high-redshift \citep{Smit2014, Tang2019, DeBarros2019, Endsley2021}.  

Within the REBELS sample, this strong correlation is present among the sample for which the dust continuum is individually detected, and it remains present for the objects with obscured SFRs determined from stacks.  This correlation is best fit by the relation $\rm \log(\rm EW([OIII]+H\beta )/ \text{\AA})=0.17\pm0.09\times \log(\rm sSFR_{\rm CSFH}/Gyr^{-1})+2.83\pm0.12$.  Errors on these parameters were determined using a bootstrap resampling method where we randomly select 40 objects from the full REBELS sample with replacement. We then perturb each chosen object by their uncertainties in sSFR and EW. This process is repeated 1000 times, and the uncertainties in the best-fit parameters are chosen to be the 16th and 84th percentiles.  When sSFR and [OIII]+H$\beta$ EW are derived using models that assume a non-parametric SFH, we achieve a similar relation of $\rm \log(\rm EW([OIII]+H\beta )/ \text{\AA})=0.41\pm0.27\times \log(\rm sSFR_{\rm Nonp}/Gyr^{-1})+2.36\pm0.28$. The nature of the non-parametric SFHs allows for an additional older stellar component that contributes significantly to the continuum flux at the wavelength of [OIII] and H$\beta$, which lowers the inferred EW (see Figure~\ref{fig:rebels12}). This additional variation leads to increased scatter in [OIII]+H$\beta$ EW and sSFR, resulting in large uncertainties in the best-fit relation between the two quantities.

%=============================================
%
%         DISCUSSION
%
%==============================================
\section{Discussion}
\label{sec:disc}
In \S4, we have presented the star forming main sequence and sSFR distribution of the REBELS sample. Here we investigate  
implications for the redshift evolution of the sSFR (\S5.1) 
and discuss why UV+IR-based sSFRs differ from those of 
SED-based measures in the REBELS sample (\S5.2). We 
close by investigating what the ALMA data reveal about 
the nature of the highest sSFR systems, in particular discussing whether the dynamical masses are consistent with the larger masses implied by non-parametric star formation histories. 

%T%he observed differences between SED-based and combined UV+FIR measurements support the physical picture where spatial variations between the dust emission and unobscured star formation is prevalent among UV-luminous objects at $z\sim7$. While this picture may be common for the most UV-luminous objects, it is not clear if these variations are present among the more compact and less luminous population. However, it is apparent that probes in the FIR, including spectroscopic redshift measurements and estimates of the dust continuum, are crucial components in understanding the sSFRs of galaxies at this epoch.  Additionally, further observations to make resolved maps of these objects is necessary to fully understand the growth of galaxies at this epoch.

\subsection{Evolution of the sSFR}
\label{sec:ssfrevo}

The average sSFR of the galaxy population provides comparison of its current stellar mass growth rate to its aggregate mass buildup. Theoretical expectations predict specific star-formation rates rise rapidly toward higher redshifts, sSFR $\propto (1+z)^{2.25}$, driven largely by the higher specific baryon accretion rates in galaxies at earlier times (e.g., \citealt{Dave2011, Dekel2009, Sparre2015}).  Observations of the sSFR evolution at high redshift have been continuously refined over the past decade \citep[e.g.,][]{Schaerer2009, McLure2011,Gonzalez2011, Stark2013, Gonzalez2014, Faisst2016, Stefanon2022} with the most recent advances coming from improved constraints on obscured star formation at high redshift from dust continuum measurements.  Most recently, ALPINE used UV and stacked FIR dust continuum measurements from ALMA to constrain the total SFRs and estimate sSFRs at high redshift. From this analysis, they reported no evolution in the sSFR at $z\simeq 4.5-5.5$ \citep{Khusanova2021},  suggesting that sSFRs may rise much less rapidly than many theoretical models predict.

\begin{figure*}
    \centering
    \includegraphics[width=1.0\linewidth]{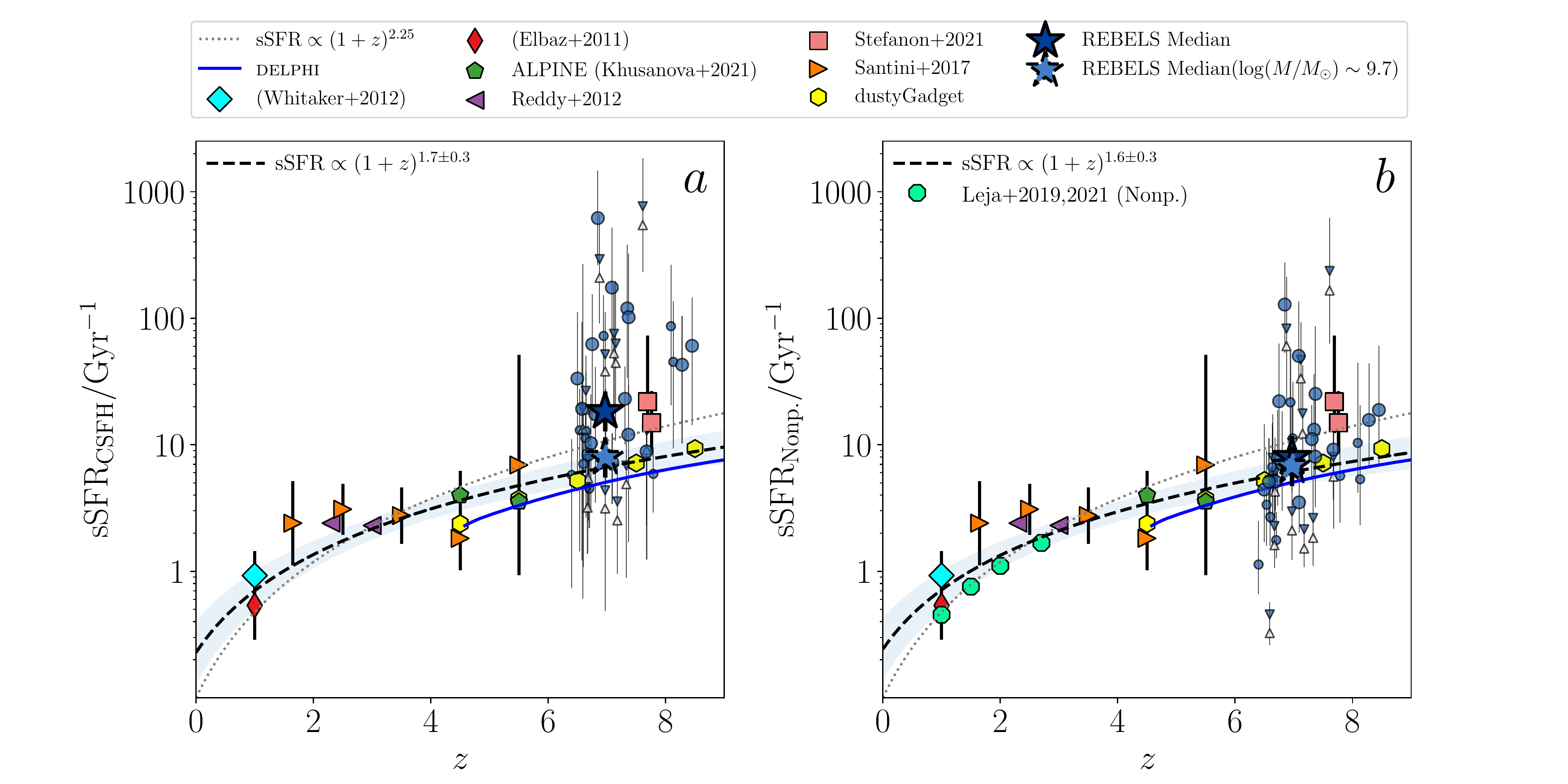}
    \caption{Observed sSFR for star-forming galaxies up to $z\sim7$.  Stellar masses for the REBELS sample were derived assuming a CSFH and using BEAGLE, and assuming a non-parametric SFH using Prospector in panels (a) and (b), respectively. At $\log(M_*/M_{\odot})\sim9.7$, the REBELS sample has a typical $\rm sSFR_{\rm{CSFH}} =8.0_{-2.3}^{+3}~\rm Gyr^{-1}$ ($\rm sSFR_{\rm Nonp} =6.2_{-1.8}^{+2.4}~\rm Gyr^{-1}$), while the full REBELS sample has $\rm sSFR_{\rm{CSFH}} =18_{-5}^{+7}~\rm Gyr^{-1}$ ($\rm sSFR_{\rm Nonp} =7.1_{-2.2}^{+2.8}~\rm Gyr^{-1}$) when masses are derived assuming a CSFH (non-parametric SFH).  This represents over a factor of two increase compared to estimates  from ALPINE at $z\sim4.5-5.5$, and an order of magnitude increase compared to at $z=1$. The best-fit power-law sSFR evolution (dashed line) is consistent with the model expectations from baryon accretion rates. The semi-analytic model, \textsc{delphi} (blue line), predicts a consistent power-law slope, however with a different overall normalization. Yellow points are taken from the median sSFR values predicted for galaxies at $z>4.4$ by recent \texttt{dustyGadget} \citep{Graziani2020} simulations of cosmic volume of $50~h^{-1}$ cMpc cube/side. The measurements from \citet{Leja2019, Leja2021} are calculated assuming a non-parametric SFH, therefore making a useful comparison to our results with the same assumption.}
    \label{fig:ssfrevolution}
\end{figure*}

The REBELS survey allows us to extend the work of ALPINE to a broader redshift range, testing for the presence of an sSFR plateau at $z> 4.5$. Figure~\ref{fig:ssfrevolution} compares the sSFRs of objects in REBELS, as well as the median of the sample, to measures at lower redshifts. For consistency to measurements at lower redshift, we focus here on sSFRs from REBELS that are derived assuming a constant star-formation history but will comment on the impact of our non-parametric models below.  The ALPINE sample consists of 118 galaxies observed in the FIR with M$_{\rm UV} < -20.2$ at $z\sim4.5-5.5$ \citep{Bethermin2020, Faisst2020, Khusanova2021}. Similar to in our analysis, the total SFRs derived for ALPINE comprise unobscured and obscured components derived from the rest-UV and IR luminosities, respectively.  However, ALPINE established IR luminosities for their sample by first deriving a relation between stellar mass and $L_{\rm IR}$ for their sample based on stacked measurements of the dust continuum. This average relation is then used to infer the obscured SFR contribution to the total SFR, and thus sSFR, of their sample \citep{Khusanova2021}. To compare our results to ALPINE we select objects from REBELS with stellar masses of $9.6 < \log(M_*/\rm M_{\odot})_{\rm CSFH} < 9.8$, which is the mass range for which the ALPINE sSFRs are established. Galaxies with stellar masses in this range have a slightly lower median redshift of $z=6.89$ compared to the full REBELS sample median of $z=6.96$. We must consider how the significant stellar mass uncertainties affect the sSFR within this mass range.  We achieve this using a bootstrap Monte-Carlo simulation, where we perturb all of the stellar masses by their uncertainties and calculate the median sSFR within the given mass window.  This process is repeated 1000 times, and the $1\sigma$ uncertainties are derived from the resulting distribution of median sSFRs.  This process results in a median $\rm sSFR_{\rm{CSFH}}=8.0_{-2.3}^{+3.0}~\rm Gyr^{-1}$ within this mass range, which is lower than the value found for the full sample.  Our estimates of the sSFR within this narrow mass range nevertheless exhibit an increase of $\times 2$ compared to the measurements from ALPINE at $z\sim4.5-5.5$, suggesting that the sSFR does increase with redshift over $4.5<z<7.0$.

A power law fit to our REBELS measurements and the ALPINE results suggests that the sSFR increases with redshift as sSFR $\propto (1+z)^{2.1\pm1.3}$ from $z\sim4.5-7.0$ (Figure~\ref{fig:ssfrevolution}).  While the uncertainties in these growth rates are large, the power law slope is consistent with the theoretical predictions 
described above \citep{Dave2011, Dekel2009, Sparre2015, Graziani2020, Pallottini2022, diCesareinprep}. We also overlay the prediction from the \textsc{delphi} semi-analytic models \citep{Dayal2014, Dayal2022} with total SFRs calculated with the same unobscured and obscured SFR conversion factors assumed for REBELS sources. These models give a consistent power-law evolution over the considered redshift range, although at slightly lower overall normalization. We finally show the values predicted by recent \texttt{dustyGadget} \citep{Graziani2020} hydrodynamical simulations of cosmic volume of 50$h^{-1}$ cMpc cube/side as yellow points. The simulations follow the assembly of dusty galaxies at $z \geq 4$ and closely reproduce the slope predicted by the Rebels sample (black dashed line) without parameter tuning. These results will be further discussed in a wider context of galaxy scaling relations at at $z \geq 4$ \citep{diCesareinprep, Grazianiinprep}. If we consider the full REBELS sample, the average stellar mass ($\log(M_*/\rm M_{\odot})=9.38$) extends to lower values than are reported in ALPINE, and the average sSFR is found to be higher ($\rm sSFR_{\rm{CSFH}} =18_{-5}^{+7}~\rm Gyr^{-1}$). These measurements suggest even more rapid sSFR evolution from $z\sim4.5$ to $z\sim7$ ($\times 4.5$) or from $z\sim5.5$ to $z\sim7$ ($\times 5$). This is notably more rapid evolution than what we found in the mass-matched sample, although this result is very sensitive to the assumed star formation history, as the low mass galaxies tend to be most impacted by the introduction of the non-parametric SFHs (\S3.3). 
Larger samples at lower masses are required across this redshift range to put the evolution implied by these higher sSFR values in context.

In the above discussion, we have limited our comparison to $z> 4.5$ with the goal of directly comparing to the ALPINE survey.  We now seek to extend our redshift baseline further.  We again adopt a fixed mass bin of $9.6 < \log(M_*/\rm M_{\odot}) < 9.8$, consistent with that adopted in ALPINE. It is crucial that these low-redshift comparison samples directly probe the obscured SFR with measures in the FIR in order to provide a self-consistent comparison to REBELS. By constraining the obscured SFRs for galaxies using {\it Spitzer}/MIPS, \citet{Whitaker2012} estimated sSFRs down to a stellar mass of $\log(M_*/M_{\odot})\sim9.5$ at $z=1$. Using this sample, they derive a $\rm sSFR =0.9_{-0.4}^{+0.5}~\rm Gyr^{-1}$ for galaxies centered at $\log(M_*/M_{\odot})=9.7$. Similarly, \citet{Elbaz2011} inferred obscured SFRs in star-forming galaxies using {\it Spitzer}/MIPS and {\it Herschel}, and found typical $\rm sSFR$ of $0.5_{-0.2}^{+0.5}~\rm Gyr^{-1}$ within the same mass range at $z\sim1$. These measurements at low redshift imply close to an order of magnitude of sSFR evolution between $z\sim1$ and $z\sim7$. When we combine these measurements with those from REBELS, we calculate a redshift evolution of the sSFR that grows as $\propto(1+z)^{1.7\pm0.3}$ over $z\sim1-7$.

As previously described, the assumed SFH can significantly affect the stellar mass, and therefore the sSFR.  While we investigated the evolution of sSFRs assuming a CSFH model for consistency with results at lower redshift, we additionally consider how the sSFR evolution would be impacted by our sample using sSFRs derived assuming a non-parametric SFH. Again for consistency with measures at lower redshift, we consider the sSFR evolution at a fixed stellar mass of $\log(M_*/M_{\odot})=9.7$.  This stellar mass is among the high-mass end of our sample, where the differences between constant and non-parametric SFHs are reduced.  As such, the difference in sSFR at this stellar mass between the two SFHs is less than for the full sample.  Assuming a non-parametric SFH, we find a value of $\rm sSFR_{\rm Nonp} = 6.2_{-1.8}^{+2.4}~\rm Gyr^{-1}$, which is only 0.11 dex lower than the median found using a CSFH in the same mass range.  As a result, we find a similar, although slightly slower redshift evolution of $\propto(1+z)^{1.6\pm0.3}$ from $z\sim1-7$ using this value. However we note that this is not a self-consistent comparison, as the lower redshift 
data have not been modeled with a similar non-parametric star formation history model. Applying such models to low-redshift galaxies has been shown to yield 0.1-0.3 dex lower sSFRs \citep{Leja2019, Leja2021}, which would imply a more rapidly rising sSFR when compared to our higher redshift datapoints (see Figure~\ref{fig:ssfrevolution}(b)).

\subsection{Implications for sSFR measurements at $z\sim7$}
\label{sec:implications}

{\it JWST} will soon deliver large samples of UV+optical SEDs, allowing the star forming main sequence to be 
calculated at a range of redshifts. In \S3.4, we 
demonstrated that within the REBELS sample, the sSFRs derived from UV+IR-based SFR 
determinations are 0.43 dex larger than those derived from the dust-corrected UV and optical SED. We have shown that the offset likely has its origin in the obscured SFR calculation, with the traditional UV+optical SED-based measurements indicating significantly lower values.

We suggest that one of the key contributing factors to 
the offset in the derived SFRs is likely to be spatial variations in the UV and FIR emission \citep[see also][]{Dayal2022, Ferrara2022}. The REBELS 
galaxies are UV-luminous systems (M$_{\rm{UV}}=-21.3$ to $-23.0$), which when viewed at high spatial resolution, tend to be composed of several star forming clumps separated by several kpc \citep{Bowler2017, Behrens2018, Sobral2019, Matthee2019, Sommovigo2020, Bowler2022, Ferrara2022, Hygate2022inprep, Inami2021prep}. These clumps are often seen to have varying levels of dust obscuration across a given galaxy \citep{Bowler2022}, leading some clumps to be brighter in the UV and others brighter in the FIR. It is important to note that in this physical picture, we attribute the nebular emission (i.e., [OIII]+H$\beta$) to the UV-dominating region, however one may expect that dust-rich regions may contribute to such emission as well \citep[see e.g.,][]{Nelson2019}. When these clumpy systems are not adequately resolved spatially (as is the norm in REBELS), the UV emission will be weighted more to the UV-bright clumps with minimal dust, leading to a blue UV slope that does not adequately capture the dust reddening experienced by more obscured clumps. This in turn will cause the dust-corrected SFR inferred from UV-optical SED fitting to be lower than the true SFR of the galaxy, similar to the offset we have found in this paper. While such spatial variations appear to be common in galaxies with similar UV luminosities as the REBELS sample, we currently do not have the required data to verify their presence in all of the REBELS systems. In the future, resolved maps of both UV and FIR emission will help shed light on this picture and its influence on the derived SFRs. Additional work will also be required to closely explore more of the systematics in the obscured SFR determination described in \S3. 

Not surprisingly given the above discussion, the UV+IR sSFRs of UV-luminous $z\sim7$ galaxies in REBELS tend to be larger than previous estimates based only on UV+optical SED fitting.  Specifically, \citet{Duncan2014} found a typical sSFR of $6.2\pm2.5 \rm ~Gyr^{-1}$ for $z\simeq 7$ galaxies with 
M$_{\rm UV} \sim-20$, nearly half of the REBELS sample median.  \citet{Stefanon2019} measured sSFRs for a sample of Lyman-break galaxies at $z\sim8$ with similar M$_{\rm UV}$ to that of REBELS and found a median sSFR of $4^{+8}_{-4}\ \rm Gyr^{-1}$, which is 0.6 dex lower than what is observed for REBELS galaxies. While further study is required to explore the origin of these differences, it seems clear that the larger obscured SFR seen in the FIR contributes significantly. Future work
will be required to investigate whether such an offset 
is also seen in the lower luminosity galaxies which will 
dominate future {\it JWST} studies.

\subsection{Properties of the highest sSFR objects}
Recent studies have established the presence of a population of very high sSFR objects at $z\gtrsim7$ \citep[e.g.,][]{Smit2015,DeBarros2019,Endsley2021, Stefanon2021}, with light dominated 
by a recent burst of star formation (i.e., few Myr). Such extreme objects have been seen to exhibit extreme [OIII]+H$\beta$ EWs \citep[i.e., $ >1000\rm \text{\AA} $ ][]{Smit2014, Smit2015, Roberts-Borsani2016, Endsley2021} 
and strong rest-UV emission lines such as CIII] and CIV \citep{Stark2015-c3, Stark2015-c4, Mainali2017, Stark2017, Laporte2017, Hutchison2019, Topping2021b}, suggestive of a significant population of young, massive stars that efficiently produce a hard ionizing spectrum \citep[e.g.,][]{Tang2019}, potentially contributing greatly 
to reionization. As 
we discussed in \S3, this population of young sources is most affected by systematics of SED modeling, 
with non-parametric star formation histories giving stellar masses that are often an order of magnitude larger than those derived from the parametric constant star formation history models. Here we discuss what ALMA measurements of [CII] and the FIR continuum reveals about this population.

Within REBELS, there are three galaxies (REBELS-09, REBELS-15 and REBELS-39) with extremely large sSFRs and very young ages based on their UV+optical  SEDs, with two of them showing [CII] emission (REBELS-15 and REBELS-39) and one showing a detection of the FIR continuum (REBELS-39). The fiducial BEAGLE models imply light-weighted CSFH ages of 1-2 Myr, and sSFRs of  120-750 Gyr$^{-1}$. These young ages are driven by the presence of large IRAC excesses which imply very large [OIII]+H$\beta$ EWs ($>$4000~\AA). The interpretation of these galaxies varies greatly with the assumed star formation history. 
The BEAGLE CSFH models suggest these are among the lowest stellar mass 
galaxies in REBELS, with derived values of  $\log(\rm M_*/M_{\odot})=8.65^{+0.35}_{-0.35}$,  $\log(\rm M_*/M_{\odot})=8.92^{+0.34}_{-0.32}$, and  $\log(\rm M_*/M_{\odot})=8.56^{+0.36}_{-0.36}$ for the three systems.  The non-parametric SFH allows these systems to have an older stellar population on top of the burst that is dominating the light, leading to larger stellar 
masses in all cases,  $\log(\rm M_*/M_{\odot})=9.20^{+0.40}_{-0.40}$,  $\log(\rm M_*/M_{\odot})=9.50^{+0.38}_{-0.40}$, and  $\log(\rm M_*/M_{\odot})=9.25^{+0.37}_{-0.32}$, respectively. The REBELS observations provide a new perspective on the gas and dust content present in these three galaxies.

We first consider whether the dynamical masses derived from [CII] provide any insight into the viability of having a significant old stellar component in REBELS galaxies with large sSFR. 
The dynamical masses will be presented in \citet{Schouws2021prep} and are calculated using the 
[CII] line widths and sizes in the rotation-dominated regime following the method described in \citet{Decarli2018}. While the derived dynamical 
masses face standard uncertainties due to the assumed velocity profile, inclination, and estimated spatial extent of [CII] emission \citep[e.g.,][]{Neeleman2021}, the typical uncertainties will not impact our primary conclusions below. 

\begin{figure*}
    \centering
    \includegraphics[width=1.0\linewidth]{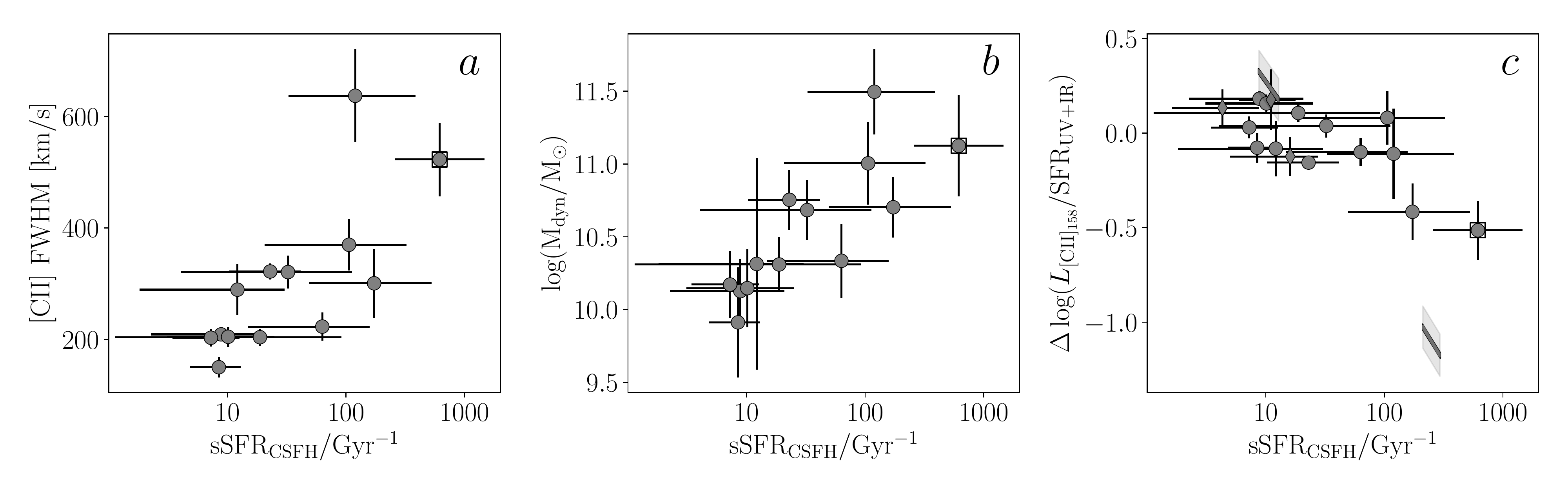}
    \caption{{\it a}: [CII] emission FWHM as a function of sSFR$_{\rm CSFH}$ for objects in REBELS with [CII] and dust continuum measurements. {\it b}: Dynamical mass calculated in the rotation-dominated regime as described in \citet{Decarli2018} plotted as a function of sSFR for REBELS galaxies detected in [CII] and the dust continuum.  {\it c}:  The excess [CII] luminosity per unit star formation relative to the relation of \citet{deLooze2014} for objects in the REBELS sample that have detections in [CII]. Objects with detections in the dust continuum are displayed as circles. The remaining objects have obscured SFRs derived from stacking as described in \S3.2.2. The diamonds indicate objects detected in the FIR stack (i.e., $\beta > -2.04$), and the grey lines indicate the range allowed by upper and lower limits on SFR$_{\rm IR}$ determined from the dust continuum stack without a detection (i.e., $\beta < -2.04$). REBELS-39 is indicated by the boxed outline.} 
  \label{fig:highestssfr}
\end{figure*}

We plot the [CII] FWHMs and dynamical masses as a function of sSFR in Figure~\ref{fig:highestssfr}(a,b). It is immediately clear in the figure that within the REBELS sample, the galaxies with the largest sSFR tend to have the largest line widths and inferred dynamical masses. 
REBELS-39 (one of the three sources discussed above and highlighted in Figure~\ref{fig:highestssfr}) provides an illustrative example. The large dynamical mass of this system ($1.3^{+1.5}_{-0.7}\times 10^{11}$ M$_\odot$) follows from 
its broad [CII] profile (FWHM=$523\pm64$ km s$^{-1}$). Clearly this is a much more massive system than is indicated by the stellar mass derived from the constant star formation modeling (3.6$\times$10$^{8}$ M$_\odot$).
The non-parametric modeling suggests a modest increase in the stellar content of REBELS-39 (1.7$\times$10$^9$ M$_\odot$), but the derived stellar mass still contributes only a small percentage of the total dynamical mass within the [CII]-emitting region of the galaxy. 
So in the case of REBELS-39, the gravitational potential can easily accommodate the presence of an older stellar population suggested by the non-parametric SFH modeling. 

A similar picture arises from the four other REBELS sources with
[CII] detections and CSFH ages below 50 Myr. The dynamical masses of
these systems are much greater than the  stellar masses implied
by non-parametric SFH modeling, with an average dynamical to stellar
mass ratio of $130$. The fact that the stellar 
mass appears to contribute such a small fraction to the dynamical
mass may suggest that these large sSFR systems have substantial
gas fractions ($>0.8$, assuming the dynamical mass is
baryon-dominated). These gas fractions 
will be discussed in more detail by \citet{Heintz2022inprep}. The presence of multiple clumps or mergers could contribute to the broad line widths and apparent large dynamical masses in these systems, due to their peculiar motions \citep[e.g.,][]{Hashimoto2019, Kohandel2019}. Indeed it is conceivable that such mergers may help contribute to the large sSFR observed in these galaxies. However, even if we adopt the dynamical masses calculated assuming the narrower line widths typical of the REBELS sample (280 km/s), we would still find values well in excess of the stellar masses derived from both the CSFH and non-parametric models.

Our main point in this paper is that the [CII] line widths of the
galaxies with the largest sSFRs point to very large dynamical masses
that can easily allow for the increase in stellar masses suggested
by non-parametric SFH modeling. While this does not confirm that
these stellar masses are correct, it does motivate further
consideration of a range of stellar masses that are possible when the
star formation history is given more flexibility. Failure to consider
these effects may lead to substantial errors in the future derivations of the stellar mass function and star forming main sequence at very high redshifts where such large sSFR systems are 
common.

We now investigate whether the [CII] output of the largest sSFR galaxies in REBELS stands out with 
respect to the majority of the sample.  As stated above, two of the three largest sSFR galaxies in REBELS 
(REBELS-15 and REBELS-39) have confident [CII] detections \citep{Schouws2021prep}. While both systems have UV and optical properties indicating large sSFR activity,
their [CII] luminosities appear much more typical of the full REBELS sample. The integrated line luminosities are $\rm L_{\rm [CII]}=1.9\pm0.4\times10^8 ~L_{\odot}$ and $7.9\pm2.5\times10^8$ L$_{\odot}$, respectively, both
very similar to the median value of detected objects
in the full REBELS sample ($\rm L_{\rm [CII]}$ of $6\pm3\times10^8 L_{\odot}$). However  
when we normalize the [CII] values by the UV+IR 
SFRs, we find that the highest sSFR systems tend to 
show a deficit with respect to the full REBELS 
sample of [CII]-detected galaxies. Figure~\ref{fig:highestssfr}(c) shows the [CII] luminosity per unit SFR relative to the predicted value from \citet{deLooze2014} as a function of sSFR.  Whereas the REBELS 
sample mostly follows the \citet{deLooze2014} relation, we can see in Figure~\ref{fig:highestssfr}(c) that REBELS-15 and REBELS-39 fall below the relation by $0.4-1.0$ dex. 
Such [CII] deficits are expected in galaxies undergoing bursts of star 
formation \citep{Ferrara2019, Pallottini2019}, 
although other physical effects 
can also contribute \citep[e.g.,][]{Croxall2012, Casey2014, Lagache2018}.

%The [CII] luminosity has been established to correlate with galaxy SFR across a range of galaxy properties in the local universe, and at high redshift \citep[e.g.,][]{deLooze2014, Schaerer2020}. 
%This link between [CII] emission and recent star formation is expected, as the massive stars produce UV radiation which is then processed by dust grains resulting in photoelectric heating of the gas.  In thermal equilibrium, the complementary cooling is provided primarily though the collisionally excited [CII] line. The exact relation between these two quantities can be modulated by a number of factors, including thermal saturation as the gas temperature increases above 92K, suppression of [CII] emission due to grain charging, and absorption of UV radiation by dust within H$\textsc{ii}$ regions. 

The dust properties of the highest sSFR galaxies are more challenging to constrain with current 
data. REBELS-39, the galaxy with the highest sSFR in our sample, shows a continuum detection with an implied  L$_{\rm IR}=4\times10^{11} \rm{L}_{\odot}$. This is a substantial IR luminosity, nearly identical to the median value of detected objects in REBELS \citep{Inami2021prep}. Such a large dust reservoir would be unexpected if we were to interpret this galaxy as among the lowest mass and youngest galaxies in the REBELS sample ($<3$Myr), as implied by the constant star formation modeling. In the context of the non-parametric model, the dust continuum could have been generated partially by the older star forming component (with age of a few 100 Myr), which dominates the stellar mass of the galaxy. We note that while the other two very high sSFR sources are not detected, their 3$\sigma$ upper limits ($<$ 4$\times$10$^{11}$ L$_\odot$) are still consistent with these systems having substantial IR luminosities. Deeper data are required to constrain the dust content of these two galaxies.

The ALMA results thus provide a new window on early galaxies with extremely large sSFR, a population of bursts which may contribute significantly to reionization. In the UV and optical, these 
objects stand out with young SEDs dominated by strong nebular 
line emission, leading to very low stellar masses 
if  constant star formation models are adopted. The [CII] line widths 
reveal that these systems are often situated in large gravitational 
potentials, with dynamical masses that can accommodate the larger 
stellar masses implied by non-parametric star formation histories. 
The [CII] and the dust continuum output are not clearly 
different from what is seen in the full REBELS galaxy sample, with luminosities occasionally  
reaching very large values. However we find that the [CII]/SFR ratio shows a significant deficit with respect to the full REBELS sample, as may be expected in systems undergoing bursts of star formation \citep[e.g.,][]{Ferrara2019, Pallottini2019, Pallottini2022}. Collectively, these results are consistent with a picture whereby the recent burst of star formation that dominates the UV and optical is just a small component within a larger galaxy. In this picture, the UV and optical is dominated by a sub-region of the galaxy that has undergone a burst, while
the ALMA observations provide a more global view of these systems, revealing large gas and dust reservoirs that may have not been expected from the UV and optical SED. This is consistent with what we suggested in \S5.2 and is similar to the resolved view seen in the first handful of UV-bright galaxies that have been observed at higher resolution \citep[e.g.,][]{Laporte2017b, Faisst2017, Bowler2017, Bowler2022}. 

%highest sSFR objects in our sample stand out in that their UV and optical properties imply a regime dominated by significant star formation, with low masses and intense ionizing radiation fields. Despite these results, these objects have ALMA properties that are not d.  This apparent decoupling between measurements in the FIR and those at shorter wavelengths points to complex spatial variations among dust and the sites of star formation being present among these UV-bright systems at $z\sim7$.  Understanding such variations will be crucial for upcoming rest-UV and optical studies from JWST.

\section{Summary}
\label{sec:summary}

In this paper we presented specific star-formation rates for a sample of 40 objects at $z\sim7-8$ observed as part of the REBELS survey.  REBELS provides a direct probe of the dust continuum in these sources, allowing improved determination of the obscured SFR. We calculate sSFRs for each galaxy, combining the derived stellar masses (from SED fitting) and SFRs (from calibrations of the UV and FIR luminosities). 

The median sSFR in the REBELS sample is sSFR$_{\rm{CSFH}}$=$18_{-5}^{+7}$ Gyr$^{-1}$ under the nominal assumptions of a constant star formation history. This value is in excess of previous estimates in the literature with similar $\rm M_{\rm UV}$ derived from SED fitting. We suggest that this offset has its origin in the obscured star formation rates, with the ALMA-based measurements giving uniformly larger values than those implied by the dust-corrected 
UV and optical SED. This effect could be explained by spatial variations in dust across individual systems, such that the components dominating the UV and optical are not always co-spatial with that dominating the 
FIR continuum. While existing data for similar systems at $z\simeq 7$ offer support for this picture \citep{Laporte2017b, Faisst2017, Bowler2017, Bowler2022}, future high spatial resolution data are required to confirm this picture for the REBELS galaxies.  

We show that the sSFRs of reionization-era galaxies are particularly sensitive to the assumed star formation history. When non-parametric 
SFHs are adopted, we find that stellar masses can increase by over an order of magnitude relative to those derived 
from constant star formation models. The changes are most significant for the youngest galaxies (e.g., $\lesssim$10 Myr) which populate the low mass end of the REBELS sample in the constant star formation models. These systems face the 
classic outshining problem, whereby the recent burst outshines the 
light from a potentially dominant earlier stellar population. We show that the dynamical masses implied by the [CII] line widths are easily able to 
accommodate the order of magnitude larger stellar masses in these young 
systems, often  suggesting these systems are capable of hosting a dominant old stellar population and 
very large gas fractions. 
While the non-parametric masses do reduce the sSFRs of the REBELS galaxies, the sample average (sSFR$_{\rm{Nonp}}$=$7.1~\rm Gyr^{-1}$) is still indicative of rapid stellar mass growth. 

Finally we characterize the redshift evolution of the sSFR for massive star forming galaxies ($9.6 < \log(\rm M_*/M_{\odot}) < 9.8$) over $1<z<7$, comparing to samples with both UV and FIR constraints on SFR. We find that the sSFR (for constant star formation models) increases with a power law that goes as $(1+z)^{1.7\pm0.3}$. Given the high mass range sampled, these results are less sensitive to the assumed star formation history, with non-parametric models at $z\simeq 7$ giving a very similar power law $(1+z)^{1.6\pm0.3}$.
In both cases, the power law increase in sSFR is only modestly shallower than the canonical power law of $(1+z)^{2.25}$ expected from evolving baryon accretion rates.

\section*{Acknowledgements}
DPS acknowledges support from the National Science Foundation
through the grant AST-2109066. RE acknowledges funding from JWST/NIRCam contract to the University of Arizona, NAS5-02015. AP acknowledges support from the ERC
Advanced Grant INTERSTELLAR H2020/740120. PD acknowledges support from the European Research Council's starting grant ERC StG-717001 (``DELPHI"), from the NWO grant 016.VIDI.189.162 (``ODIN") and the European Commission's and University of Groningen's CO-FUND Rosalind Franklin program. RB acknowledges support from an STFC Ernest Rutherford Fellowship [grant number ST/T003596/1]. YF acknowledge support from NAOJ ALMA Scientific Research Grant number 2020-16B.

\bibliographystyle{apj}
\bibliography{REBELS_sSFR}

\end{document}